\documentclass[a4paper,11pt]{article}
\pdfoutput=1

\usepackage{jcappub}
\usepackage[T1]{fontenc}
\usepackage{physics}
\usepackage{booktabs}
\usepackage{array}
\usepackage{float}
\newcommand{\PreserveBackslash}[1]{\let\temp=\\#1\let\\=\temp}
\newcolumntype{C}[1]{>{\PreserveBackslash\centering}p{#1}}
\newcolumntype{R}[1]{>{\PreserveBackslash\raggedleft}p{#1}}
\newcolumntype{L}[1]{>{\PreserveBackslash\raggedright}p{#1}}

\newcommand{\dphi}{\delta\phi}
\newcommand{\Pphi}{\mathcal{P}_\phi}
\newcommand{\Pphis}{\mathcal{P}_{\phi,\sigma}}
\newcommand{\Pphiseff}{\tilde{\mathcal{P}}_{\phi,\sigma}}
\newcommand{\Ppi}{\mathcal{P}_\pi}
\newcommand{\Pphipi}{\mathcal{P}_{\phi\pi}}
\newcommand{\R}{\mathcal{R}}
\newcommand{\PR}{\mathcal{P}_\mathcal{R}}

\newcommand{\PReff}{\tilde{\mathcal{P}}_\mathcal{R}}
\newcommand{\PRseff}{\tilde{\mathcal{P}}_{\mathcal{R},\sigma}}
\newcommand{\Pe}{\tilde{P}}
\newcommand{\eps}{\epsilon}
\newcommand{\epseff}{\tilde{\epsilon}}
\newcommand{\xivec}{{\hat{\xi}}}
\newcommand{\xii}{\hat{\xi}_i}
\newcommand{\mxii}{\bar{\xi}_i}
\newcommand{\mxivec}{\bar{\xi}}
\newcommand{\dxii}{\delta\xi_i}
\newcommand{\dxij}{\delta\xi_j}
\newcommand{\dxivec}{\delta\xi}
\newcommand{\xij}{\hat{\xi}_j}
\newcommand{\xiij}{\hat{\xi}_{i,j}}
\newcommand{\kc}{k_\sigma}
\newcommand{\phieff}{\tilde{\phi}}
\newcommand{\Neff}{\tilde{N}}
\newcommand{\ini}{\text{ini}}
\newcommand{\fin}{\text{c}}
\newcommand{\out}{\text{final}}
\newcommand{\sumi}{\sum_{i=1}^n}
\newcommand{\bpartial}{\bar{\partial}}
\newcommand{\Eeff}{\tilde{E}}
\newcommand{\CC}{C\nolinebreak\hspace{-.05em}\raisebox{.4ex}{\tiny\bf +}\nolinebreak\hspace{-.10em}\raisebox{.4ex}{\tiny\bf +}}
\def\CC{{C\nolinebreak[4]\hspace{-.05em}\raisebox{.4ex}{\tiny\bf ++}}}

\title{\boldmath Numerical stochastic inflation constrained by frozen noise}

\author{Eemeli Tomberg}
\affiliation{Laboratory of High Energy and Computational Physics, National Institute of Chemical Physics and Biophysics, R{\"a}vala pst.~10, Tallinn, 10143, Estonia}
\emailAdd{eemeli.tomberg@kbfi.ee}

\abstract{
Stochastic inflation can resolve strong inflationary perturbations, which seed primordial black holes. I present a fast and accurate way to compute these perturbations in typical black hole producing single-field models, treating the short-wavelength Fourier modes beyond the de Sitter approximation. The squeezing and freezing of the modes reduces the problem to one dimension, and the resulting new form of the stochastic equations, dubbed `constrained stochastic inflation,' can be solved efficiently with semi-analytical techniques and numerical importance sampling. In an example case, the perturbation distribution is resolved in seconds deep into its non-Gaussian tail, a speed-up of factor $10^9$ compared to a previous study. Along the way, I comment on the role of the momentum constraint in stochastic inflation.
}

\begin{document}
\maketitle
\flushbottom

\section{Introduction}

Cosmic inflation \cite{Starobinsky:1980te,Kazanas:1980tx,Guth:1980zm} sources cosmological perturbations originating from the quantum vacuum \cite{Lyth:2009zz}. Typical perturbations behave linearly and follow Gaussian statistics, completely described by the perturbation power spectrum. A linear analysis is enough to capture, for example, most of the statistical properties of the cosmic microwave background radiation (CMB) \cite{Planck:2018jri}. However, rare, strong perturbations behave non-linearly. They form primordial black holes (PBHs) \cite{Carr:1974nx,Carr:1975qj}, a dark matter candidate \cite{Chapline:1975ojl,Carr:2020gox,Carr:2020xqk,Green:2020jor,Carr:2021bzv}, and a possible source of gravitational waves \cite{Raidal:2017mfl,Ali-Haimoud:2017rtz}. To accurately predict the PBH abundance in a given model of inflation, one has to employ non-linear techniques to compute the probability distribution of inflationary perturbations.

Stochastic inflation \cite{Starobinsky:1986fx, Morikawa:1989xz, Salopek:1990jq, Salopek:1990re, Habib:1992ci, Starobinsky:1994bd, Mijic:1994vv, Matacz:1996fv, Bellini:1996uh, Woodard:2005cw, Martin:2005hb, Tsamis:2005hd, Martin:2005ir, Kunze:2006tu, vanderMeulen:2007ah, Finelli:2008zg, Beneke:2012kn, Gautier:2013aoa, Fujita:2013cna, Garbrecht:2013coa, PerreaultLevasseur:2013eno, PerreaultLevasseur:2013kfq, Garbrecht:2014dca, Fujita:2014tja, PerreaultLevasseur:2014ziv, Burgess:2015ajz, Vennin:2015hra, Boyanovsky:2015jen, Onemli:2015pma, Boyanovsky:2015tba, Moss:2016uix, Grain:2017dqa, Tokuda:2017fdh, Collins:2017haz, Pattison:2017mbe, Prokopec:2017vxx, Cruces:2018cvq, Pinol:2018euk, Ezquiaga:2018gbw, Biagetti:2018pjj, Choudhury:2018rjl, Firouzjahi:2018vet, Ezquiaga:2019ftu, Pattison:2019hef, Markkanen:2019kpv, Prokopec:2019srf, Hardwick:2019uex, Pinol:2020cdp, Cable:2020dke, Ando:2020fjm, De:2020hdo, Bounakis:2020jdx, Figueroa:2020jkf, Firouzjahi:2020jrj, Ballesteros:2020sre, Cohen:2021fzf, Cruces:2021iwq, Andersen:2021lii, Rigopoulos:2021nhv, Pattison:2021oen, Achucarro:2021pdh, Kamenshchik:2021tjh, Tomberg:2021xxv, Figueroa:2021zah, Tada:2021zzj, Cruces:2022imf, Ahmadi:2022lsm, Mahbub:2022osb, Animali:2022otk, Jackson:2022unc, Cable:2022uwd} provides such a technique. There, one coarse-grains the system over super-Hubble scales and keeps track of the coarse-grained local background and the short-wavelength perturbations separately. The coarse-grained quantities follow the non-linear classical FLRW equations and receive stochastic kicks from the quantum mechanical short-wavelength fluctuations. This approximates the full quantum gravity calculation, aiming to include the leading non-linearites while keeping the computation tractable.

Originally, stochastic inflation was studied in slow-roll (SR) inflation, and the short-wavelength perturbations were taken to have their de Sitter form, where the field fluctuation is proportional to the Hubble parameter \cite{Starobinsky:1986fx}. Recently, it has been used to study models with abundant PBH production \cite{Pattison:2017mbe, Cruces:2018cvq, Ezquiaga:2018gbw, Biagetti:2018pjj, Firouzjahi:2018vet, Ezquiaga:2019ftu, Pattison:2019hef, Prokopec:2019srf, Ando:2020fjm, De:2020hdo, Figueroa:2020jkf, Ballesteros:2020sre, Cruces:2021iwq, Rigopoulos:2021nhv, Pattison:2021oen, Achucarro:2021pdh, Hooshangi:2021ubn, Tomberg:2021xxv, Figueroa:2021zah, Tada:2021zzj, Cruces:2022imf, Ahmadi:2022lsm, Animali:2022otk, Jackson:2022unc}. The simplest models include a single canonical scalar field whose potential has a feature---a flat section or a low local maximum. As the field rolls over the feature, the slow-roll approximation is broken, and the perturbations grow, leading to a high PBH abundance. Most of these studies still assume de Sitter perturbations, and many use further approximations e.g.~to the inflaton potential to obtain analytical results, providing a quick way to estimate PBH statistics. Such studies can be instructive in understanding the general features of the perturbation probability distribution. They have shown that the Gaussian approximation indeed breaks for large enough perturbations, and the distribution's tail can be best described with a simple exponential, or a sum thereof \cite{Pattison:2017mbe, Ezquiaga:2019ftu, Prokopec:2019srf, Rigopoulos:2021nhv, Pattison:2021oen, Achucarro:2021pdh, Tada:2021zzj, Ahmadi:2022lsm, Animali:2022otk}.

At the same time, simplifying assumptions can compromise the accuracy of the computation, leading to large uncertainties in the PBH estimates. The full, unsimplified problem is best attacked numerically \cite{Ezquiaga:2019ftu, De:2020hdo, Figueroa:2020jkf, Pattison:2021oen, Tomberg:2021xxv, Figueroa:2021zah, Animali:2022otk, Jackson:2022unc}. In particular, \cite{Figueroa:2020jkf, Tomberg:2021xxv, Figueroa:2021zah} dropped the de Sitter assumption and evolved the short-wavelength perturbations alongside the coarse-grained field in a generic single-field potential, presenting the most careful study of the full stochastic system to date. While accurate, such computations are numerically expensive and are not suitable for scans over a parameter space or studying vast collections of models. They also lack the transparency of the analytical results.

On a related note, a particular line of inquiry into stochastic inflation emphasizes the importance of the momentum constraint, one of the components of Einstein's equation \cite{Salopek:1990jq, Salopek:1990re, Prokopec:2019srf, Cruces:2021iwq, Rigopoulos:2021nhv, Cruces:2022imf}. This restricts the stochastic equations and may impact PBH predictions. However, there is no consensus in the community on the significance of this constraint for stochastic inflation.

In this paper, I take the detailed numerical computation of \cite{Figueroa:2020jkf, Figueroa:2021zah} and reproduce the results with semi-analytical techniques using minimal computational resources. I show that, regardless of any fundamental considerations, the momentum constraint is followed in practice for perturbation modes that are frozen and squeezed when they reach the coarse-graining scale. In a typical PBH-producing single-field model, this applies to all the important modes. I use this to reformulate the stochastic equations into a new form, dubbed here \emph{constrained stochastic inflation}. I solve the new equations in two example models both with an analytical approximation and numerically using importance sampling, a technique first considered for stochastic inflation in \cite{Jackson:2022unc}. All computations are done with the full perturbation power spectrum instead of the de Sitter approximation. The results match those of the numerically expensive computation of \cite{Figueroa:2020jkf, Figueroa:2021zah}. The presented method to compute the perturbation statistics is thus fast, practical, and transparent, but also accurate, combining the benefits of analytical and numerical studies.

The paper is organized as follows: Section~\ref{sec:stoch} introduces the stochastic formalism following the setup of \cite{Figueroa:2020jkf, Figueroa:2021zah}, presents the two example models, and examines the momentum constraint. Section~\ref{sec:new_formalism} establishes the constrained stochastic equations and considers their solution analytically. Section~\ref{sec:numerics} presents numerical solutions to these equations and compares them to previous results. Section~\ref{sec:discussion} is reserved for discussion, and section~\ref{sec:conclusions} concludes the paper.

\section{Stochastic inflation}
\label{sec:stoch}
I study canonical single-field models of inflation, with the action
\begin{equation} \label{eq:S}
    S = \int \dd^4 x \sqrt{-g}\qty[\frac{1}{2} R - \frac{1}{2}\partial^\mu\varphi\partial_\mu\varphi - V(\varphi)] \, ,
\end{equation}
where $R$ is the Ricci scalar, $\varphi$ is the inflaton and $V$ is its potential, and I set the reduced Planck mass to one. As is standard, I work in the perturbated FRLW universe and divide $\varphi$ into long and short wavelength parts, separated by the \emph{coarse-graining scale} $\kc$:
\begin{equation} \label{eq:phi_division}
\begin{split}
    \varphi(N,\vec{x}) &\equiv
    \phi(N,\vec{x}) + \dphi(N,\vec{x}) \\
    &= \int_{k<\kc} \frac{\dd^3 k}{(2\pi)^{2/3}}\phi_k(N) e^{-i\vec{k} \cdot \vec{x}} +
    \int_{k>\kc} \frac{\dd^3 k}{(2\pi)^{2/3}}\dphi_k(N) e^{-i\vec{k} \cdot \vec{x}} \, .
\end{split}
\end{equation}
Here $\phi_k$ and $\dphi_k$ both refer to Fourier modes of the total field $\varphi$, but have been renamed for easier bookkeeping. The time variable is the number of e-folds of expansion of space, $N = \ln a$, where $a$ is the FLRW scale factor.\footnote{Using $N$ as a time variable is practical for reasons related to the gauge choice of cosmological perturbations and for the ease of use of the $\Delta N$ formalism, discussed below.} 
The coarse-graining scale $\kc$ is a function of time, defined as $k \equiv \sigma a H$, with $\sigma$ a constant and $H$ the Hubble parameter. Choosing $\sigma < 1$ places the coarse-graining at super-Hubble scales at all times.

Such a choice of coarse-graining scale has two consequences: first, the long-wavelength part $\phi$ is approximately constant in one super-Hubble patch and its spatial derivatives can be neglected there; and second, Fourier modes constantly drift across the coarse-graining scale, leaving the short-wavelength regime and joining the averaged long-wavelength field $\phi$. The Einstein equations in one super-Hubble patch can then be approximated as (see e.g.~\cite{Figueroa:2021zah})
\begin{gather}
    \label{eq:bg_eom}
    \phi' = \pi + \xi_\phi \, ,
    \qquad
    \pi' = -\qty(3 - \eps_1)\pi - \frac{V'(\phi)}{H^2} + \xi_\pi \, ,
    \qquad
    H^2 = \frac{V(\phi)}{3-\eps_1} \, ,
    \\
    \label{eq:pert_eom}
    \dphi_k'' = -\qty(3 - \eps_1)\dphi_k' - \qty[\frac{k^2}{a^2H^2} +
    2\eps_1\qty(3-\eps_1) +
    2\pi\frac{V'(\phi)}{H^2} + \frac{V''(\phi)}{H^2}
    ]\dphi_k \, ,
\end{gather}
where $\pi$ is the momentum associated to $\phi$, and I introduced the first slow-roll parameter $\eps_1 = \frac{1}{2}\pi^2$ to shorten the notation. Throughout the paper, a prime denotes a derivative w.r.t. $N$, except if the function to be differentiated has an explicit argument, in which case a prime denotes a derivative w.r.t. this argument.

The local background equations \eqref{eq:bg_eom} are of the standard FLRW form, except for the added noise terms $\xi_\phi$ and $\xi_\pi$ introduced by the drifting Fourier modes. They introduce stochastic kicks to the `classical' field evolution at every time step. The word `classical' is used throughout the paper to refer to evolution without stochastic noise. I will also use the special notation $\bpartial_N$ to denote a classical $N$-derivative without the $\xi$-terms, that is, $\bpartial_N\phi \equiv \pi$, $\bpartial_N\pi \equiv -(3-\eps_1) - V'(\phi)/H^2$.

The noise originates from the short wavelength perturbations that are random due to their quantum origin. I treat the short-wavelength modes linearly in the spatially flat gauge;\footnote{Since we have chosen $N$ as the time variable, it must not receive stochastic kicks from the short-wavelength perturbations. In principle, this means we should work in the uniform-$N$ gauge. However, working in the spatially flat gauge is technically simpler, and the two gauges are practically identical in the super-Hubble limit, as shown analytically in \cite{Pattison:2019hef} and numerically in \cite{Figueroa:2021zah}.} \eqref{eq:pert_eom} is the corresponding version of the Sasaki--Mukhanov equation. The linear modes start from the Bunch--Davies vacuum with
\begin{equation} \label{eq:Bunch-Davies}
    \delta \phi_k = \frac{1}{\sqrt{2k}a} \, , \qquad \delta (a\phi_k)' = -i\frac{k}{H}\dphi_k \, , \qquad k \gg aH \, ,
\end{equation}
and follow Gaussian statistics, inducing Gaussian noise with the two-point correlators \cite{Pattison:2019hef, Figueroa:2020jkf, Figueroa:2021zah}
\begin{subequations}\label{eq:noise_correlators}
\begin{align}
    \label{eq:noise_correlator_phiphi}
    \expval{\xi_\phi(N)\xi_\phi(N')} &= \frac{1}{6\pi^2}\frac{\dd \kc^3}{\dd N}|\dphi_{\kc}(N)|^2\delta(N-N') =
    \qty(1-\eps_1)\Pphi(N,\kc)\delta(N-N') \, , \\
    \label{eq:noise_correlator_pipi}
    \expval{\xi_\pi(N)\xi_\pi(N')} &= \frac{1}{6\pi^2}\frac{\dd \kc^3}{\dd N}|\dphi'_{\kc}(N)|^2\delta(N-N') =
    \qty(1-\eps_1)\Ppi(N,\kc)\delta(N-N') \, , \\
    \label{eq:noise_correlator_phipi}
    \expval{\xi_\phi(N)\xi_\pi(N')} &= \frac{1}{6\pi^2}\frac{\dd \kc^3}{\dd N}\dphi_{\kc}(N)\dphi'^*_{\kc}(N)\delta(N-N') =
    \qty(1-\eps_1)\Pphipi(N,\kc)\delta(N-N') \, ,
\end{align}
\end{subequations}
where we recognized $\frac{k^3}{2\pi^2}|\dphi_k(N)|^2\equiv\Pphi(N,k)$, the power spectrum of the field perturbations for a given wavenumber. Similarly, I defined $\frac{k^3}{2\pi^2}|\dphi'_k(N)|^2\equiv\Ppi(N,k)$ and $\frac{k^3}{2\pi^2}\dphi_k(N)\dphi'^*_k(N)\equiv\Pphipi(N,k)$. Later, I will use the short-hand notation $\mathcal{P}_{X,\sigma}(N) \equiv \mathcal{P}_{X}(N,\kc(N))$ to denote the power spectrum at the wavenumber of the coarse-graining scale at time $N$.

Equations \eqref{eq:bg_eom}--\eqref{eq:noise_correlators} are the starting point for the stochastic computations in this paper; for a longer discussion on their derivation, see e.g. \cite{Pattison:2019hef, Figueroa:2021zah}. To get a handle on cosmological perturbations, one may use the $\Delta N$ formalism \cite{Sasaki:1995aw, Sasaki:1998ug, Wands:2000dp, Lyth:2004gb}: the amount of local expansion $N$ is related to the coarse-grained curvature perturbation $\psi_c$ by 
\begin{equation} \label{eq:delta_N_def}
    \Delta N \equiv N - \expval{N} = \psi_c \, ,
\end{equation}
where $\expval{N}$ is the unperturbed mean e-fold number. Following the procedure of \cite{Figueroa:2020jkf, Figueroa:2021zah}, one solves the equations \eqref{eq:bg_eom}--\eqref{eq:pert_eom} for $\phi$ and a range of modes $\dphi_k$ for many realizations of the stochastic noise, corresponding to many super-Hubble patches of space, starting from an unperturbed hypersurface at early times and ending at a hypersurface with a fixed $\phi=\phi_\text{final}$. The stochastic noise is turned off in the middle of this evolution when a desired final coarse-graining scale $\kc=k_c$ is reached; for an approximately constant $H$, this happens at a fixed $N=N_c$.\footnote{This procedure differs somewhat from the popular first passage time method \cite{Vennin:2015hra}. I will discuss the differences in section~\ref{sec:discussion}.} The rest of the evolution up to $\phi_\text{final}$ is computed without stochastic kicks. This way, all the patches will have a fixed comoving size $\sim 1/k_c$ at the final constant-$\varphi$ hypersurface, and their $\Delta N$ values all correspond to curvature perturbations in the comoving gauge coarse-grained over $1/k_c$, $\psi_c = \R_c$. One then builds the probability distribution $p(\R_c)$ from the sample points. Since $\R$ freezes at super-Hubble scales, the obtained distribution will maintain its shape until the corresponding scales re-enter the Hubble radius after the end of inflation.

In linear perturbation theory, $p(\R_c)$ would be Gaussian. However, the local background equations \eqref{eq:bg_eom} are highly non-linear through $V$, $H$, and $\eps_1$. In addition, the $\dphi_k$ modes affect the evolution of $\phi$ through the noise terms, and $\phi$ affects $\dphi_k$ in turn through the background dependency in \eqref{eq:pert_eom}, introducing a non-linear backreaction loop. In practice though, it was shown in \cite{Figueroa:2021zah} for various example models that the backreaction is not important: it is enough to solve equation \eqref{eq:pert_eom} once for each $k$ in a classical background and use the resulting $\dphi_k$ modes to source the noise at a given time. I will explain this behavior in section~\ref{sec:efold_matching}. Even then, the non-linearities modify $p(\R_c)$ for large perturbations, giving it a non-Gaussian tail at large $\R_c$ \cite{Pattison:2017mbe, Ezquiaga:2019ftu, Prokopec:2019srf, Figueroa:2020jkf, Rigopoulos:2021nhv, Pattison:2021oen, Achucarro:2021pdh, Tada:2021zzj, Ahmadi:2022lsm, Figueroa:2021zah, Animali:2022otk, Jackson:2022unc}.

The inclusion of non-linearities is the chief merit of stochastic inflation and makes it the tool of choice when computing e.g. the statistics of PBHs, which form from strong perturbations with $\R_c \sim 1$ \cite{Harada:2013epa, Escriva:2019phb}. However, it is still an approximative method, aiming to include the most important non-linearities of the system while keeping the problem computationally feasible. One point of approximation is related to the choice of the $\sigma$ parameter. A large $\sigma$ pushes more modes into the long-wavelength regime and thus includes more of the non-linear interactions. However, if $\sigma$ is too large, the gradient approximation that enabled us to neglect the spatial derivatives in equation~\eqref{eq:bg_eom} no longer applies. In the examples below, I use $\sigma = 0.01$; I will return to the $\sigma$-dependency of the results in section~\ref{sec:discussion}.

\subsection{Example models and background evolution}
\label{sec:background}

\begin{figure}
    \centering
    \includegraphics{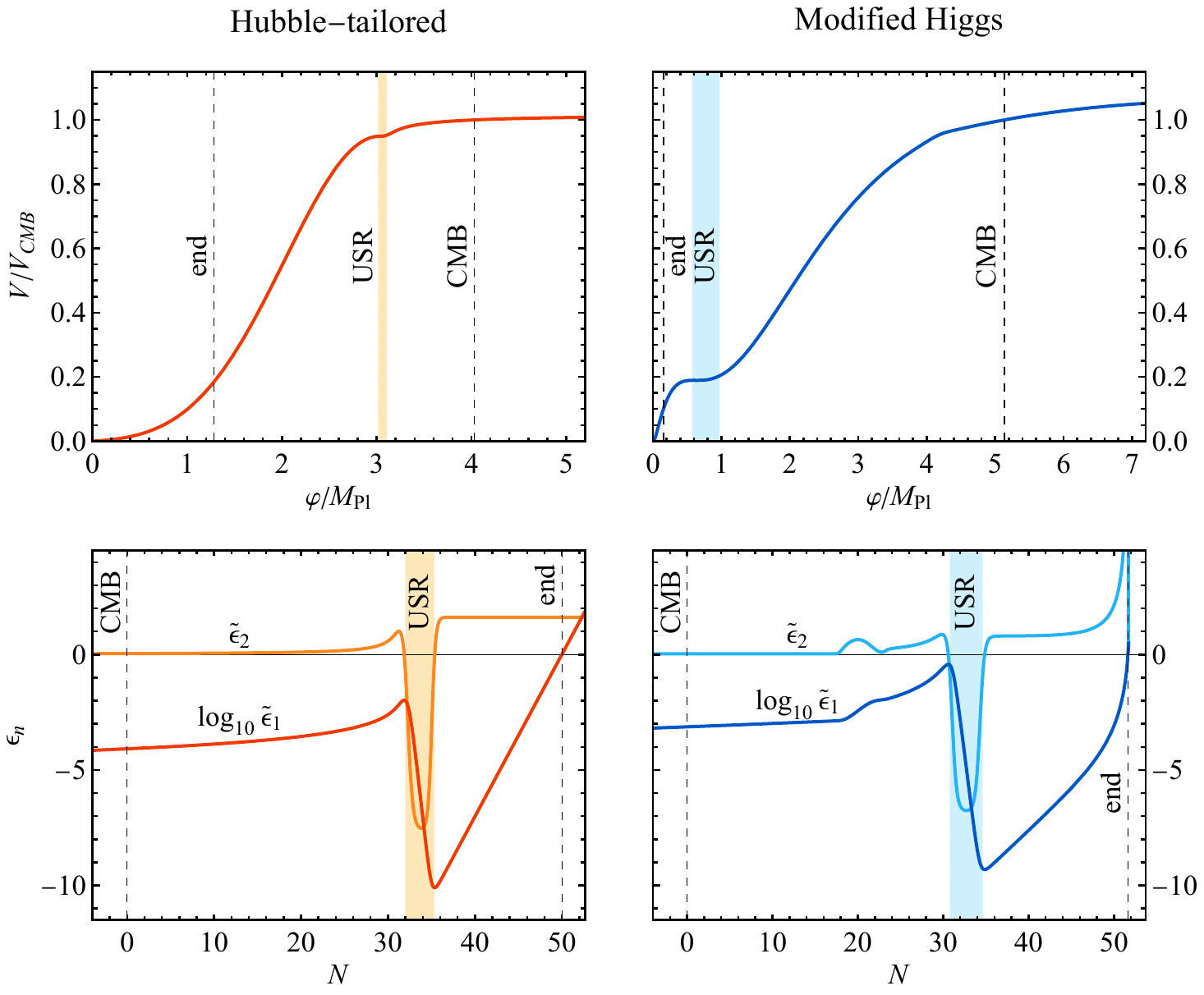}
    \caption{The potential $V$ and the two slow-roll parameters $\eps_1$, $\eps_2$ in the Hubble-tailored and modified Higgs models. `CMB' indicates the field and e-fold values where the CMB pivot scale $0.05 \, \text{Mpc}^{-1}$ exits the Hubble radius, `USR' indicates the ultra-slow-roll period, and `end' indicates the end of inflation. The potentials are normalized to their CMB values. Both have a local maximum in the USR regime, accompanied by a local minimum at slightly larger field values, though these are not easy to see by eye in the figure. The e-folds $N$ are computed from the CMB scale onwards.}
    \label{fig:Vepsgrid}
\end{figure}

In this paper, I consider two PBH-producing models of inflation. The first of these is the `Asteroid mass' potential from \cite{Figueroa:2020jkf, Figueroa:2021zah}, called here the modified Higgs model due to its origin as a hand-tuned version of Higgs inflation with a running Higgs self-coupling \cite{Rasanen:2018fom}. The second I call the Hubble-tailored model, built semi-analytically from the e-fold dependence of the Hubble and slow-roll parameters to produce perturbations with a tunable strength, as explained in appendix~\ref{sec:eff_model}. The basic properties of these models are depicted in figure~\ref{fig:Vepsgrid}.

The inflaton potentials of both models have a long plateau at large field values and a strong feature with a local maximum at small field values. During its evolution, the inflaton starts from the plateau, rolls down and over the feature, and ends up in the minimum at $\varphi=0$ where the universe reheats. The evolution can be described in terms of the slow-roll variables
\begin{equation} \label{eq:SR_parameters}
    \eps_1 \equiv -\bpartial_N \ln H = \frac{1}{2}\pi^2 \, , \qquad
    \eps_2 \equiv \bpartial_N \ln \eps_1 = 2\frac{\bpartial_N\pi}{\pi} \, ,
\end{equation}
so that $\eps_1 < 1$ is a sufficient and necessary condition for inflation, whereas slow-roll (SR) inflation also requires $|\eps_2| \lesssim 1$. Note the use of the noiseless derivative $\bpartial_N$. As the field rolls over the feature in the potential, $\eps_2$ dips to highly negative values, $\eps_2 \lesssim -6$, during a period of ultra-slow-roll (USR) inflation. In the figures of this paper, the highlighted USR region corresponds to $\eps_2 < -2$ to be consistent with the convention of \cite{Figueroa:2020jkf, Figueroa:2021zah}. The USR period is important for amplifying cosmological perturbations.
After USR, when the field is still close to the local potential maximum, there's a period of constant-roll (CR) inflation where $\eps_2$ is a positive constant that can be larger than one, connected to the USR period by the Wands duality \cite{Wands:1998yp}. For a recent exploration of these different phases, see \cite{Karam:2022nym}.

The above discussion describes the field evolution in the absence of stochastic noise. To differentiate this classical trajectory from a general stochastic trajectory $\phi(N)$, I will indicate the field value on the classical trajectory by $\phieff$ and the corresponding e-fold number by $\Neff$. Since classical field evolution is monotonic, it specifies a one-to-one mapping $\phieff \mapsto \Neff$ between field and e-fold values. We will exploit this feature in section~\ref{sec:new_formalism}. I will use a similar notation for other quantities on the classical trajectory, such as the first slow-roll parameter $\epseff_1$.

\subsection{Perturbation evolution and power spectrum}
\label{sec:perturbations}

Inflationary perturbations are typically discussed in terms of the comoving curvature perturbation $\R$. It is related to the field perturbation $\dphi$ by \cite{Lyth:2009zz}
\begin{equation} \label{eq:R}
    \R = \frac{\dphi}{\pi} = \frac{\dphi}{\sqrt{2\eps_1}} \, .
\end{equation}
CMB observations constrain perturbations around the pivot scale $k_* = 0.05 \, \text{Mpc}^{-1}$ \cite{Planck:2018jri,BICEP:2021xfz} as
\begin{equation} \label{eq:CMB}
\begin{gathered}
    A_s = \PR(k_*) = \frac{H^2}{8\pi^2\eps_1} \approx 2.1\times10^{-9} \, , \\
    \quad n_s = 1 - \eps_2 - 2\eps_1 = 0.9649 \pm 0.0042\, , \qquad r = 16\eps_1 < 0.036 \, .
\end{gathered}
\end{equation}
Here $\PR(k)=\frac{k^3}{2\pi^2}|\R_k|^2$ is the power spectrum of $\R$, $n_s$ is its running, and $r$ is the tensor-to-scalar ratio. The expressions in terms of the slow-roll parameters apply in the SR limit and should be evaluated at the Hubble exit of $k_*$. The example models of this paper produce CMB predictions compatible with the observations.

Below the CMB scale, $\PR$ is not strongly constrained. Our models of interest have a peak in $\PR$ produced by USR where $\eps_1$ in \eqref{eq:R} is small. It is these strong perturbations that lead to abundant PBH formation. From the stochastic point of view, a high $\PR$ makes the noise terms in \eqref{eq:bg_eom} strong, as we will see explicitly below. The SR approximation does not apply for these modes, and the perturbations have to be solved numerically from \eqref{eq:pert_eom}.

\begin{figure}
    \centering
    \includegraphics{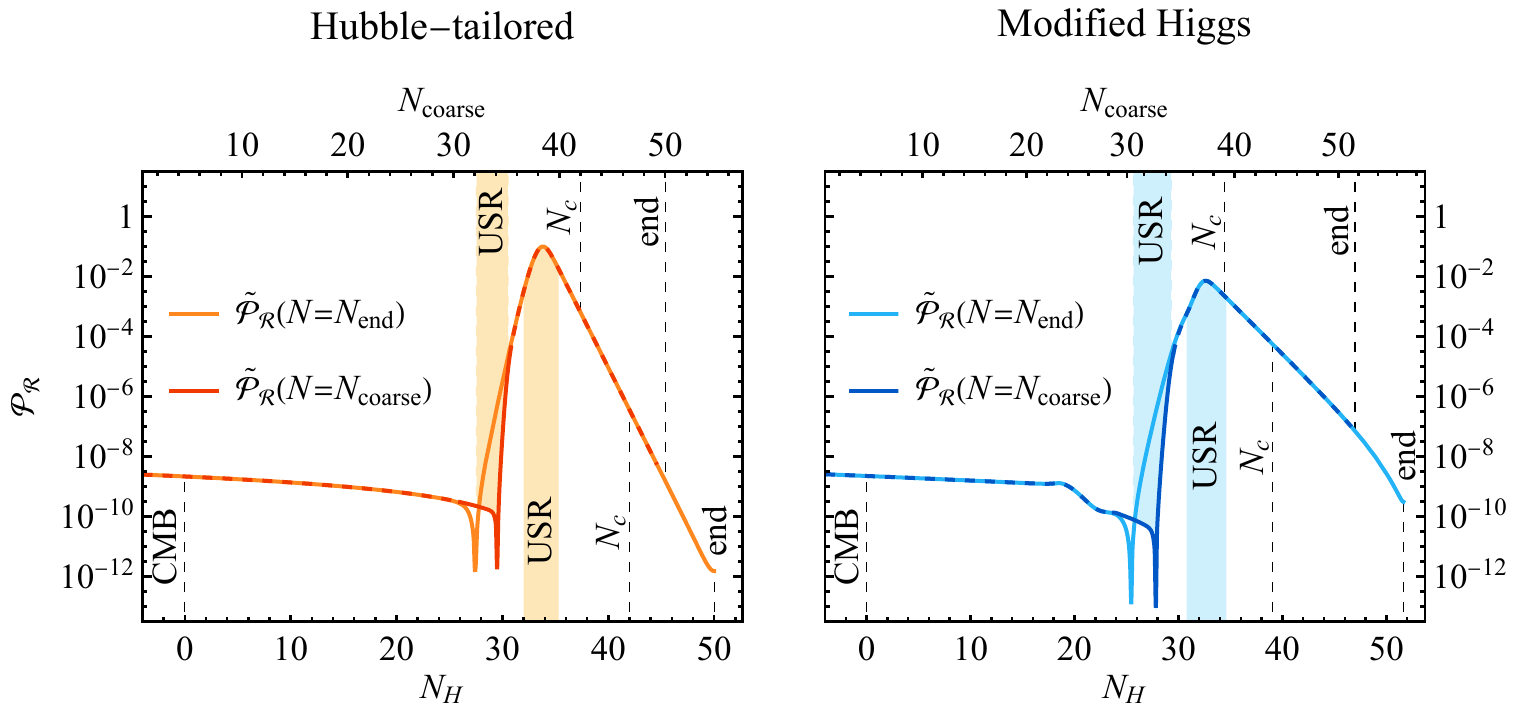}
    \caption{Power spectra $\PR$ in the two models, evaluated at two different times: at the end of inflation (when all modes are frozen) and at the time when the mode crosses the coarse-graining scale. The e-fold numbers label the modes: $N_H$ is the Hubble exit of the mode, and $N_\text{coarse}$ indicates when it crossed the coarse-graining scale with the chosen coarse-graining parameter $\sigma=0.01$. Different times are highlighted on the $N$-axes identically to figure~\ref{fig:Vepsgrid}, with the addition of $N_c$, the time when the final mode that contributes a kick crosses the coarse-graining scale (see the discussion below equation~\eqref{eq:delta_n_vs_c}).}
    \label{fig:PRgrid}
\end{figure}

Figure~\ref{fig:PRgrid} shows $\PR$ in the example models computed in the classical background. Consistently with the notation of section \ref{sec:background}, I will denote these power spectra by $\PReff$ to emphasize that the backreaction between the background and the perturbations in \eqref{eq:bg_eom}--\eqref{eq:pert_eom} has been neglected in their computation.

At this point, two properties of the perturbations in the super-Hubble limit $k \ll aH$ need to be pointed out. Their importance for the stochastic formalism will become evident in the next section.

\begin{figure}
    \centering
    \includegraphics{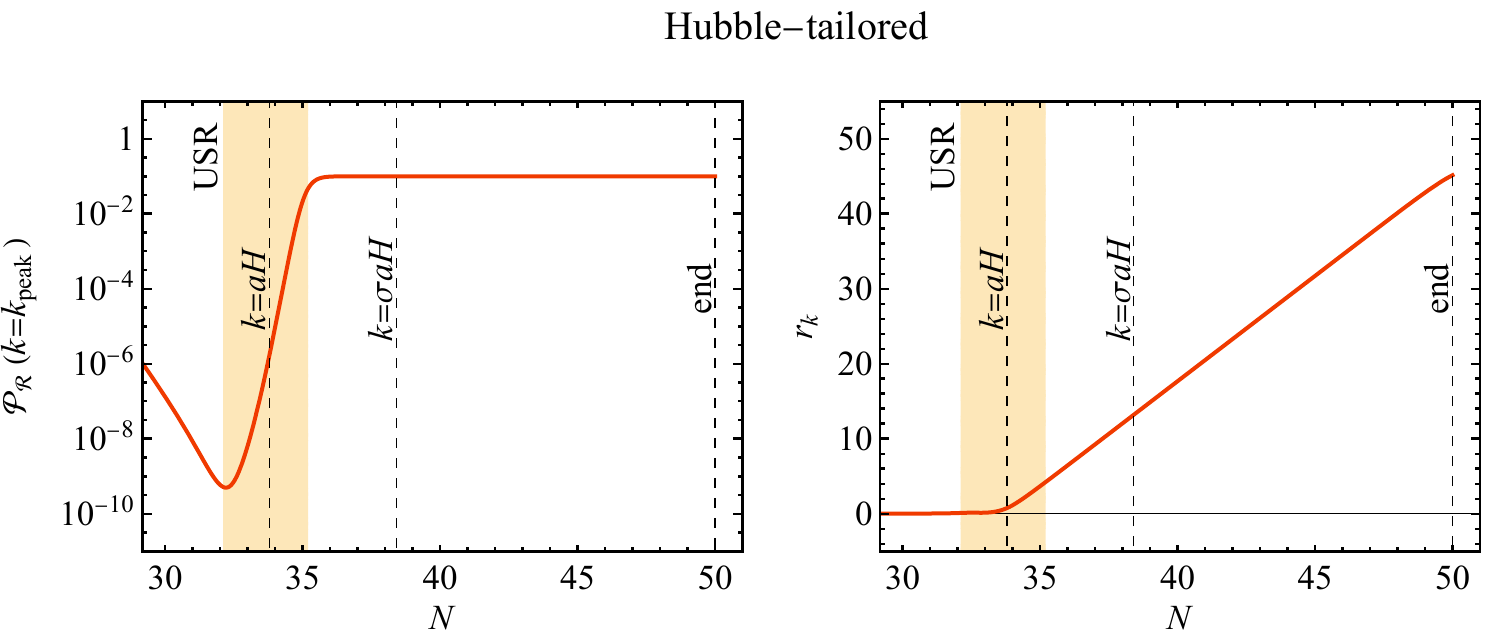}
    \caption{Time evolution of the power spectrum $\PR(k)$ and squeezing $r_k$ of the Fourier mode corresponding to the power spectrum peak, $k_\text{peak} \approx 2.3 \times 10^{13} \, \text{Mpc}^{-1}$, in the Hubble-tailored model. The Hubble exit $k=aH$ and the coarse-graining time $k=\sigma aH$ are highlighted. The mode freezes soon after the end of USR. Squeezing increases linearly from the Hubble exit.}
    \label{fig:example_mode}
\end{figure}

\paragraph{Freezing.} In the super-Hubble limit, in the classical background, the perturbation equation~\eqref{eq:pert_eom} can be written in terms of $\R$ in the simple form
\begin{equation} \label{eq:R_eom}
    \R_k'' + \qty(3 - \eps_1 + \eps_2)\R'_k = 0 \, ,
\end{equation}
and it has the general solution
\begin{equation} \label{eq:R_solution}
    \R_k = A_k + B_k \int \frac{\dd N}{a^3 H \eps_1} \, ,
\end{equation}
consisting of a constant and a dynamical term. Equation~\eqref{eq:R_eom} implies that $\R_k$ approaches the constant solution if $\eps_2 > -3 + \eps_1$, but grows when $\eps_2 < -3 + \eps_1$. In the first case, the $B_k$-term in \eqref{eq:R_solution} decays, whereas in the second case, it grows. In particular, in the initial SR phase, $\R_k$ approaches a constant: the curvature perturbation \emph{freezes} after Hubble exit \cite{Lyth:2009zz}. In the USR phase, with $\eps_1 \ll 1$ and a strongly negative $\eps_2$, the curvature perturbation may grow exponentially, leading to the well-known super-Hubble enhancement of $\R$. However, in the final CR phase, the freezing behavior takes over again. The power spectra of figure~\ref{fig:PRgrid} show $\PR$ both at the end of inflation at this frozen value and at the coarse-graining time. We see that most modes have reached the final, frozen value by the time of coarse-graining; the exception are modes exiting the Hubble radius near the beginning of USR. Figure~\ref{fig:example_mode} shows the time evolution of $\PR$ for an example mode, demonstrating the freezing.

\paragraph{Squeezing.} The quantum perturbations of the field and its momentum are correlated, as can be seen in \eqref{eq:noise_correlator_phipi}. The level of correlation is described by the squeezing coefficient $r_k$ \cite{Grishchuk:1990bj, Albrecht:1992kf, Martin:2007bw, Martin:2012ua, Grain:2019vnq}, which in our case can be written as \cite{Figueroa:2021zah}
\begin{equation} \label{eq:squeezing}
    \cosh (2 r_k) = k a^2\qty(|\dphi_k|^2 + \frac{a^2H^2}{k^2}|\dphi'_k|^2) \, .
\end{equation}
Figure \ref{fig:example_mode} shows the evolution of $r_k$ for an example mode. In the initial Bunch--Davies vacuum, $r_k = 0$, with both terms in \eqref{eq:squeezing} contributing equally to the hyperbolic cosine. After Hubble exit, the second term comes to dominate and $r_k$ starts to grow. In \cite{Figueroa:2021zah}, $r_k$ was reported to momentarily decrease during USR---this is possible close to the Hubble exit when mode behavior is non-trivial, but eventually, all modes get highly squeezed simply due to \eqref{eq:squeezing} growing with $a$. In particular, when the mode has frozen during the CR phase, we have $\dphi \sim \sqrt{\eps_1}$ from \eqref{eq:R}, so that $\dphi' \sim \sqrt{\eps_1}\eps_2$, and with $\eps_2\sim\text{const.} > 0$ and $\eps_1 \sim a^{\eps_2}$ we get $\cosh(2r_k) \sim a^{4+\eps_2}$. This constant growth rate is actually achieved already before freezing, as can be seen in figure~\ref{fig:example_mode}; this is due to the Wands duality, which dictates that $\dphi_k$ grows with the same rate throughout the USR and CR phases \cite{Wands:1998yp, Karam:2022nym}.

Since high squeezing implies a high correlation between the field and momentum perturbations, the noises $\xi_\phi$ and $\xi_\pi$ are not independent. Instead, one determines the other, according to\footnote{In the squeezed limit, $\frac{\dphi'_k}{\dphi_k}$ is real \cite{Figueroa:2021zah}, so \eqref{eq:noise_correlators} is consistent with real-valued noise.}
\begin{equation} \label{eq:noise_squeezed_correlation}
    \xi_\pi = \xi_\phi\frac{\dphi'_k}{\dphi_k}\bigg|_{k=\kc} \, .
\end{equation}
This relation is consistent with the correlation functions \eqref{eq:noise_correlators}, but it is a stronger statement: not only the two-point functions of $\xi_\pi$ and $\xi_\phi$ but also the realized noises are related so that in each step of the stochastic evolution, the noise arises from only one independent Gaussian random variable instead of two.


In our example models, all important modes close to the peak of $\PR$ are frozen and highly squeezed by the time they reach the coarse-graining scale. This statement depends on the parameter $\sigma$, which sets the delay between the Hubble exit and the coarse-graining of a mode. However, even our moderately large value of $\sigma=0.01$ is enough to guarantee this behavior quite generically. During USR, the power spectrum grows roughly as $k^4$ \cite{Byrnes:2018txb, Carrilho:2019oqg, Ozsoy:2019lyy, Tasinato:2020vdk, Karam:2022nym}. If we wish $\PR$ to grow at this rate from its CMB value of $2.1\times 10^{-9}$ to a maximum of, say, $\PR\sim 0.1$, USR can last at most $4.4$ e-folds. A longer USR phase leads to too strong perturbations with dominant stochastic effects, which overproduce PBHs \cite{Rigopoulos:2021nhv}. In practice, USR tends to be shorter, $3.0$ e-folds in our Hubble-tailored model and $3.5$ e-folds in the modified Higgs model. In comparison, modes exiting the Hubble radius at the beginning of USR get coarse-grained $\ln 100 \approx 4.6$ e-folds later. Thus, the strong USR modes get coarse-grained only after the end of USR, in the subsequent CR phase, when they are frozen and squeezed.

\subsection{Role of the momentum constraint}
\label{sec:constraint}

A series of papers \cite{Prokopec:2019srf, Cruces:2021iwq, Rigopoulos:2021nhv, Cruces:2022imf} has promoted the importance of the \emph{momentum constraint} for stochastic inflation. It is an additional equation for the local background quantities on top of \eqref{eq:bg_eom}, arising from time-space components of the Einstein equations, describing spatial derivatives. In the notation of \cite{Prokopec:2019srf}, using the ADM formalism, the constraint reads
\begin{equation} \label{eq:momentum_constraint}
    \bar{K}^j_{\ i|j} - \frac{2}{3}K_{|i} + \Pi\phi_{|i}=0 \, ,
\end{equation}
where $\bar{K}^i_{\,j}$ is the traceless part of the extrinsic curvature tensor, $K$ is the trace,
$\Pi$ is the field momentum (defined differently from $\pi$ in \eqref{eq:bg_eom}), a vertical bar indicates the covariant derivative on a spatial slice, and the $i,j$ indices refer to spatial directions.

In the long-wavelength limit where spatial derivatives are sub-leading, the  Einstein equations give
\begin{equation} \label{eq:ext_curv_vanishing}
    \bar{K}^i_{\,j} \sim e^{-3\alpha} \, ,
\end{equation}
where $\alpha$ measures the expansion of space, essentially the local number of e-folds $N$. In other words, $\bar{K}^i_{\,j}$ quickly vanishes at super-Hubble scales, erased by the expansion. Once it is gone, \eqref{eq:momentum_constraint} becomes
\begin{equation} \label{eq:homog_momentum_constraint}
    \partial_i H = -\frac{1}{2}\dot{\phi}\partial_i \phi \, ,
\end{equation}
where we used the long-wavelength correspondences $K\to-3H$ and
$\Pi\to\dot{\phi}$, and a dot indicates a derivative w.r.t.~the cosmic time.

Equation~\eqref{eq:homog_momentum_constraint} connects the different super-Hubble patches together. Using it with the other components of the Einstein equations, one can show \cite{Salopek:1990jq, Prokopec:2019srf} that $H$ and $\phi$ must be in a one-to-one relationship, $H=H(\phi)$. This constrains the evolution of the local background on one fixed trajectory. The behavior is easy to understand in slow-roll inflation with an attractor trajectory, which all solutions approach. However, the momentum constraint suggests that this is true beyond slow-roll: over time, in some sense, all trajectories converge to one. In \cite{Prokopec:2019srf, Cruces:2021iwq, Rigopoulos:2021nhv, Cruces:2022imf}, this was used to simplify the stochastic formalism. Enforcing \eqref{eq:homog_momentum_constraint} means that the field and its momentum follow a fixed path in phase space, with stochastic kicks moving the field back and forth on this path but never outside of it.

I next point out a loophole in the reasoning of \cite{Prokopec:2019srf} and show that the fixed-trajectory behavior does not, in fact, need to apply in stochastic inflation. This was already noted in \cite{Salopek:1990jq, Salopek:1990re}. The crux of the matter lies in equation~\eqref{eq:ext_curv_vanishing}: while $\bar{K}^i_{\,j}$ vanishes dynamically over time, it is not forced to be identically zero by any fundamental considerations. Equation~\eqref{eq:homog_momentum_constraint}, and the following fixed-trajectory behavior, only applies when enough time has passed for $\bar{K}^i_{\,j}$ to decay. Importantly, in stochastic inflation, $\bar{K}^i_{\,j}$ has a new source of time evolution beyond this decay: it is sourced by quantum fluctuations emerging from the vacuum, an effect absent in a computation based on classical general relativity. The stochastic noise can momentarily increase $\bar{K}^i_{\,j}$ before the classical behavior erases it again. In the presence of such noise, \eqref{eq:homog_momentum_constraint} is broken. The full momentum constraint \eqref{eq:momentum_constraint} is still satisfied: its role is to give $\bar{K}^i_{\,j}$ in a way that always stitches the different patches together in a consistent manner. Integrated over long distances, a small $\bar{K}^i_{\,j}$ can lead to large differences between the trajectories of two far-away patches. A small $\bar{K}^i_{\,j}$ can still be neglected locally inside one patch; the patches evolve independently (this is called the \emph{separate universe approach}, see e.g. \cite{Salopek:1990jq, Wands:2000dp}).

Even if \eqref{eq:homog_momentum_constraint} does not apply on a fundamental level, it may still be valid phenomenologically under specific circumstances. In fact, the super-Hubble freezing of the curvature perturbations accomplishes exactly this: a decaying mode dies away, corresponding to the decay of $\bar{K}^i_{\,j}$ in \eqref{eq:ext_curv_vanishing}, and the stochastic evolution gets confined on a fixed trajectory. To see how this happens, I write the time derivative of the definition \eqref{eq:R} in the suggestive form
\begin{equation} \label{eq:perturbation_direction}
    \frac{\dphi_k'}{\dphi_k} = \frac{\bpartial_N \pi}{\bpartial_N \phi} + \frac{\R_k'}{\R_k} \, .
\end{equation}
In the frozen limit $\R_k'/\R_k \to 0$, the ratio of the momentum and field perturbations equals the ratio of the classical momentum and field time derivatives. Since frozen perturbations are squeezed as well, \eqref{eq:noise_squeezed_correlation} applies, showing that the ratio of the momentum and field kicks is exactly aligned with the classical evolution: if the field is kicked by $\dd\phi=\dd N\bpartial_N\phi$ (where $\dd N$ is a small constant), then the momentum is kicked by $\dd\pi=\dd N\bpartial_N\pi$, and the system simply moves along its classical trajectory by the e-fold jump $\dd N$. In fact, this result is not surprising: in the super-Hubble limit, the perturbation equation \eqref{eq:pert_eom} is just the linearized form of the background equations \eqref{eq:bg_eom}, and hence it is solved by the difference between two nearby classical solutions, that is, by $\dphi_k = \dd N\bpartial_N\phi$ for a constant $\dd N$. I have shown that when the perturbations freeze, this solution becomes an attractor, akin to $\bar{K}^i_{\,j}$ dying out in \eqref{eq:ext_curv_vanishing}.

In the previous section, we saw that in PBH-producing models of single-field inflation, the strongest perturbations are quite generically frozen by the time they contribute their stochastic kicks after the end of the USR period. Thus \eqref{eq:homog_momentum_constraint} and the fixed-trajectory behavior applies. The dynamics at the time of the stochastic kicks is more important than the dynamics at the Hubble exit of a given mode. Freezing is related to the attractor behaviour of the CR phase that follows USR: all super-Hubble patches fall onto the same attractor background trajectory, described by a single clock variable, and non-adiabatic perturbations in perpendicular directions die out.
However, caution is in order: if the leading stochastic noise was applied during USR, say, as a consequence of a large coarse-graining parameter $\sigma$, then there is no attractor, the exiting modes may not be frozen, and the fixed-trajectory behavior may break down.

\section{Constrained stochastic formalism}
\label{sec:new_formalism}

I will now use the lessons learned in the previous sections to reformulate the stochastic inflation formalism. Before that, it is convenient to move from the continuum equations \eqref{eq:bg_eom}--\eqref{eq:pert_eom} to discrete time steps of length $\dd N$, corresponding to discrete steps $\dd \phi$ of the field. The field noise from \eqref{eq:bg_eom} becomes
\begin{equation} \label{eq:noise_discretization}
    \xi_\phi \to \sqrt{(1-\eps_1)\Pphis/\dd N} \, \xii \, ,
\end{equation}
where I used the short-hand $\Pphis$ for the power spectrum of the mode currently giving a kick, defined below \eqref{eq:noise_correlators}. I separated the noise amplitude of \eqref{eq:noise_correlator_phiphi} from the normally distributed random variables $\xii$. These are independent and have unit variance, that is,
\begin{equation} \label{xii_correlators}
    \expval{\xii\xij} = \delta_{ij} \, .
\end{equation}
The indices $i,j$ enumerate the time steps.
The $1/\sqrt{\dd N}$ factor in \eqref{eq:noise_discretization} produces the correct continuum limit for the correlators.

I then restrict movement onto the fixed classical trajectory $\phieff(\Neff)$ from section~\ref{sec:background}. This simplifies the stochastic equations considerably. First, the canonical variables $\phi$ and $\pi$ are no longer independent, so the two equations for $\phi$ and $\pi$ in \eqref{eq:bg_eom} are condensed into one,
\begin{equation} \label{eq:discrete_phi_eom}
    \frac{\dd \phi}{\dd N} = \pi + \sqrt{\qty(1-\eps_1)\Pphis/\dd N} \, \xii \, .
\end{equation}
The classical drift $\pi$, corresponding to the field velocity in the absence of noise, is given by $\phieff'(\Neff)$. In other words, $\pi$ is evaluated on the classical trajectory at the current $\phi$.\footnote{I emphasize that $\pi$ in \eqref{eq:discrete_phi_eom} is a function of $\phi$ only; this function must be solved separately by solving the classical trajectory. As $\phi$ undergoes stochastic motion, so does $\pi=\pi(\phi)$.} Similarly, we have $\eps_1 \to \epseff_1 = \frac{1}{2}\phieff'^2(\Neff)$ in the noise coefficient. Second, instead of $\phi$, we can use $\Neff$ as the stochastic variable that indicates the position on the trajectory. We then solve $\Neff(N)$ instead of $\phi(N)$, with the substitution $\phi(N) = \phieff(\Neff(N))$, giving $\dd \phi = \phieff'(\Neff) \dd \Neff$. Moving the terms in \eqref{eq:discrete_phi_eom} around gives
\begin{equation} \label{eq:discrete_neff_eom}
    {\dd \Neff} = \dd N + \sqrt{\qty[1-\epseff_1(\Neff)]\frac{\Pphis}{2\epseff_1(\Neff)} \dd N} \, \xii \, .
\end{equation}
This gives the change of $\Neff$ in one time step $\dd N$. I have written the time arguments out explicitly to clarify the functional dependencies on the stochastic variable $\Neff$ versus the clock time and actual amount of spatial expansion $N$.

Ambiguity still remains in the evaluation of $\Pphis$. I use the result of \cite{Figueroa:2021zah} discussed in section~\ref{sec:stoch}, according to which the power spectrum can be pre-computed in a noiseless background and the result can be used to give the norm of the noise in the stochastic equations. In other words, I take $\Pphis \to \Pphiseff$ in accordance with the convention of section~\ref{sec:perturbations}. We then need to determine the scale $\kc$ that contributes to the noise at a particular time $N$. Since the stochastic and classical background evolutions are different, there is some ambiguity in this matching. Two natural options arise. We can write $\Pphiseff = \Pphiseff(\Neff)$, matching the classical and stochastic evolutions by their field values: $\Neff$ gives a field value $\phi=\phieff(\Neff)$, and we evaluate the classical power spectrum at this field value. Alternatively, we can take $\Pphiseff = \Pphiseff(N)$, doing the matching directly through the time variable. I will consider both options below.

To solve the stochastic equation \eqref{eq:discrete_neff_eom}, I fix the initial condition $\Neff_\ini = N_\ini$ at some early time before the occurrence of the power spectrum peak. The exact starting point does not matter; as long as the kicks are small there, the effect on the final $\Neff$ is negligible. As described in section~\ref{sec:stoch}, we then evolve the system forward until the time $N_\fin$, corresponding to some $\Neff_\fin$ that depends on the realization of the stochastic noise. With its discrete time steps, equation \eqref{eq:discrete_neff_eom} is suitable as-is for numerical solving, and I will do this in section~\ref{sec:numerics}. With a fixed $\dd N$, it takes a fixed number of time steps, denoted below by $n$, to reach $N_\fin$. To obtain $\Delta N$, we would normally continue to evolve the system non-stochastically after $N_\fin$ until a hypersurface with a fixed field value $\phi_\out$ is reached. This corresponds to a fixed $\Neff_\out$ but a stochastically varying $N_\out$, with $\expval{N_\out} = \Neff_\out$ and thus $\Delta N = N_\out - \Neff_\out$ from \eqref{eq:delta_N_def}. However, since $\Neff$ and $N$ evolve in sync along the classical trajectory when $\xi_\phi=0$, we have $N_\out - N_\fin = \Neff_\out - \Neff_\fin$, and thus we can simply write
\begin{equation} \label{eq:delta_n_from_neff}
    \Delta N = N_\fin - \Neff_\fin \, .
\end{equation}
We can even go one step further and define $\Delta N = N - \Neff$ moment-by-moment during the stochastic evolution. This allows us to keep track of the time evolution of $\Delta N$ from zero toward its final value, and we can see which scales contribute the most to the final result. In addition, this makes it easy to numerically compute the curvature perturbations coarse-grained over multiple scales: perform a number of stochastic simulations, store $N - \Neff$ at multiple time steps in each simulation, and build the statistics for each of these different scales. However, in the numerical examples below, I will concentrate on one time scale with a fixed $N_\fin$ and stick to the definition \eqref{eq:delta_n_from_neff}.

\subsection{Gaussian limit: independent kicks}
\label{sec:Gaussian_limit}
Let us briefly examine the limit of small perturbations, $\Delta N \ll 1$, where $\Neff$ only deviates slightly from $N$. We can then write \eqref{eq:discrete_neff_eom} as
\begin{equation} \label{eq:discrete_neff_eom_gaussian}
    {\dd \Neff} = \dd N + \sqrt{\Pe(N) \dd N} \, \xii \, ,
    \qquad
    \Pe(N) \equiv \qty[1-\epseff_1(N)]\PRseff(N) \, ,
\end{equation}
where I used \eqref{eq:R} to write $\Pphi/(2\eps) = \PR$. Since the right-hand side is independent of $\Neff$, no memory of the previous evolution is preserved, and all time steps contribute to $\Delta N$ independently. The
$\Delta N$ distribution can then be easily integrated. As a sum of independent Gaussian random variables, $\Delta N$ is itself Gaussian, and its variance is the sum of the components' variances:
\begin{equation} \label{eq:delta_n_variance_gaussian}
    \expval{\Delta N^2}
    = \sumi \Pe(N_i) \dd N
    \xrightarrow{\dd N \to 0}
    \int_{N_\ini}^{N_\fin} \Pe(N) \dd N
    \approx
    \int_{k_\ini}^{k_\fin} \PReff(k) \, \dd \ln k \, .
\end{equation}
In the last step, I assumed $\epseff_1 \ll 1$ and hence $\dd N \approx \dd \ln k$, and dropped the time dependence of $\PR$, assuming it has reached its final frozen value by the time the modes give their kicks. We then recover the standard result of $\expval{\Delta N^2}$ expressed as an integral over the curvature power spectrum. This was used in e.g.~\cite{Fujita:2013cna, Fujita:2014tja, Vennin:2015hra, Pattison:2017mbe} to compute the power spectrum through stochastic methods. The current formulation makes the origin of this result transparent.

In typical inflationary scenarios, we expect the stochastic corrections to be small; then the $\Delta N$ distribution is indeed Gaussian near its peak, and \eqref{eq:delta_n_variance_gaussian} is a good approximation for its width. This was demonstrated numerically for multiple example models in \cite{Figueroa:2020jkf, Figueroa:2021zah}. Only farther in the tail of the probability distribution does the approximation $\Neff \approx N$ fail and non-Gaussianities start to accumulate.

\subsection{Field value matched perturbations}
\label{sec:field_matching}

Let us now consider the field value matched perturbations, $\Pphis = \Pphiseff(\Neff)$. The stochastic equation becomes
\begin{equation} \label{eq:discrete_neff_eom_field_matched}
    {\dd \Neff} = \dd N + \sqrt{\Pe(\Neff) \dd N} \, \xii \, .
\end{equation}
Compared to the Gaussian case, the $\Neff$-dependence of the right-hand side introduces memory effects and complicates the analysis. On the other hand, there is now no explicit $N$ dependence, which allows us to still make some progress analytically.

The stochastic nature of $\xii$ ensures that each realization of $\Neff(N)$ is different. However, for each $\Delta N$, there is a `most probable' path around which the realizations cluster. I treat the noises $\xii$ as components of a $n$-dimensional vector $\xivec$, and write the probability density in this vector space as
\begin{equation} \label{eq:p_xi}
    p(\xivec) = \frac{1}{(2\pi)^{n/2}} \exp(-\frac{1}{2}|\xivec|^2) \, , \qquad |\xivec|^2 \equiv \sumi \xii^2 \, .
\end{equation}
Using \eqref{eq:discrete_neff_eom_field_matched}, I write the exponent as
\begin{equation} \label{eq:S_xi}
    S_\xi \equiv -\frac{1}{2} |\xivec|^2
    = -\sumi \frac{(\Neff'-1)^2}{2\Pe(\Neff)}\dd N
    \xrightarrow{\dd N \to 0}
    -\int_{N_\ini}^{N_\fin} \frac{(\Neff'-1)^2}{2\Pe(\Neff)} \dd N \, .
\end{equation}
Note that the continuum limit behaves well and the dependence on the $N$ step length vanishes. The most probable paths minimize $S_\xi$. It is essentially an action integral for $\Neff$; studying the probability distribution around the most probable paths is akin to the saddle point approximation of a path integral. Varying $S_\xi$ with respect to $\Neff(N)$ gives the Euler--Lagrange equation
\begin{equation} \label{eq:neff_path_field_match}
    \Neff'' + \frac{\Pe'(\Neff)}{2\Pe(\Neff)}\qty(1 - \Neff'^2) = 0 \, ,
\end{equation}
which can be integrated to give the simpler form
\begin{equation} \label{eq:dneff_path_field_match}
    \Neff' = \sqrt{1-c\Pe(\Neff)} \, ,
\end{equation}
where $c$ is an integration constant, analogous to the conserved energy of a mechanical system with no explicit time dependence.

\begin{figure}
    \centering
    \includegraphics{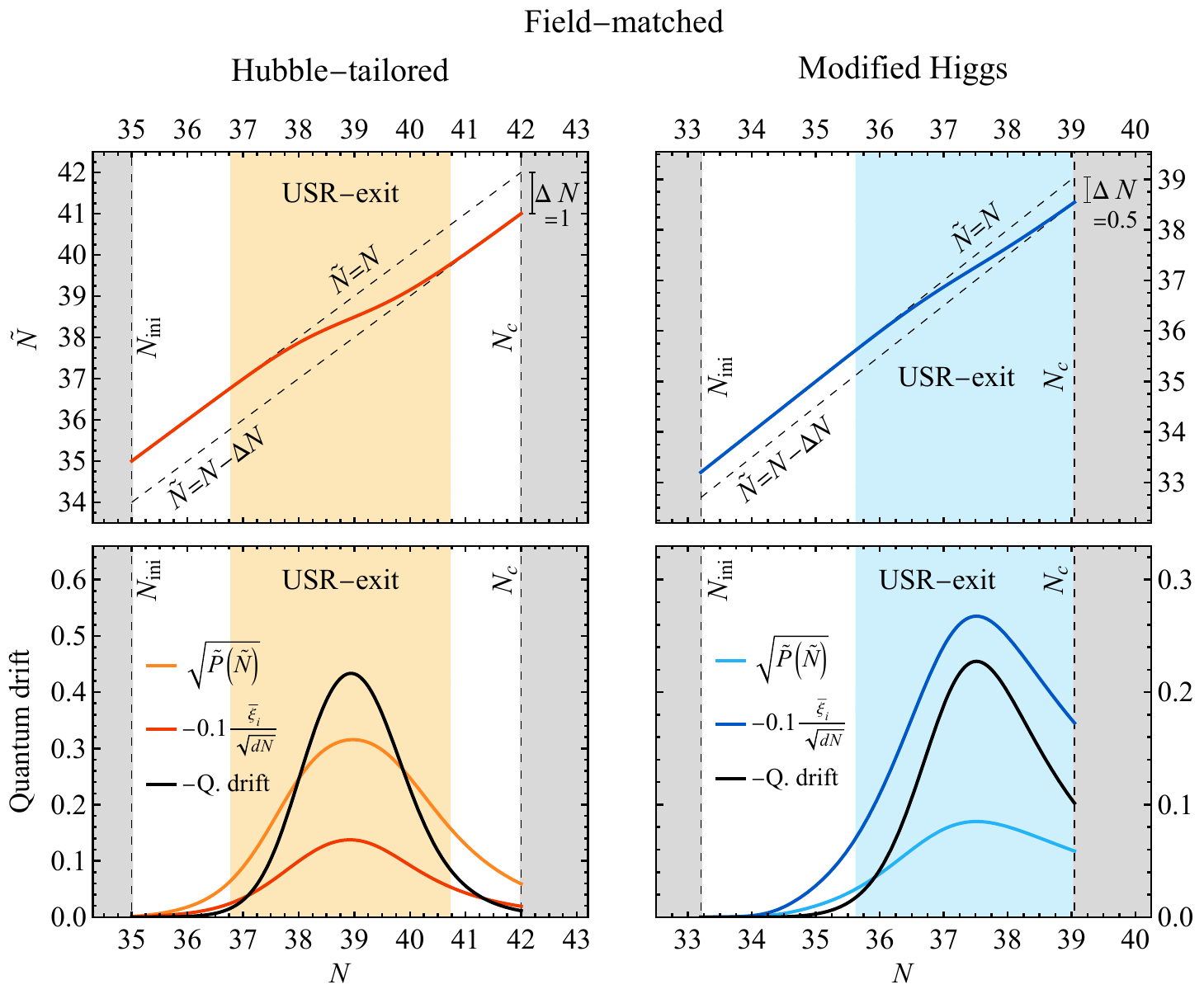}
    \caption{Examples of the most probable paths $\Neff(N)$ in the field-matched case, solved from \eqref{eq:dneff_path_field_match}, for both of the example models. The colored band `USR-exit' indicates when the strong modes that exited the Hubble radius during USR are giving their kicks. Around this time, $\Neff'$ differs from its classical value of one. The lower panels show the components that make up the total quantum drift on these paths, that is, the noise contribution to $\Neff'$ given by $\sqrt{\Pe(\Neff)/\dd N}\mxii$. In both cases, the quantum drift stays below the classical one.}
    \label{fig:field_matching}
\end{figure}

When solving \eqref{eq:dneff_path_field_match}, the initial $\Neff$ is fixed to $N_\ini$ as discussed above, but different $c$ values correspond to different initial conditions for $\Neff'$ and a different final $\Delta N$ from \eqref{eq:delta_n_from_neff}. We can immediately see that $c=0$ corresponds to $\Neff=N$, or $\Delta N = 0$, and $c > 0$ ($c < 0$) corresponds to $\Neff' < 1$ ($\Neff' > 1$) and thus $\Delta N > 0$ ($\Delta N < 0$). To clarify the connection, let us write
\begin{equation} \label{eq:delta_n_vs_c}
    \Delta N = \int_{N_\ini}^{\Neff_\fin} \qty(\frac{\dd N}{\dd \Neff} - 1) \dd \Neff = \int_{N_\ini}^{N_\fin-\Delta N} \qty(\frac{1}{\sqrt{1-c\Pe(\Neff)}} - 1) \dd \Neff \, .
\end{equation}
From here, we can numerically match a $c$ to a $\Delta N$. The full $\Neff(N)$ path can then be integrated from \eqref{eq:dneff_path_field_match}, and examples of this are shown in figure~\ref{fig:field_matching}. We see that $\Neff' \approx 1$ in the beginning and in the end; the contribution to $\Delta N$ arises from the large-perturbation scales in the middle, where both $\Pe$ and $\xii$ peak.

To compute the probability density of $\Delta N$, I convert \eqref{eq:p_xi} into the $\Delta N$ space as
\begin{equation} \label{eq:p_to_dn}
    p(\Delta N)\dd (\Delta N) =
    \hspace{-1em}
    \int\displaylimits_{D(\Delta N)}
    \hspace{-0.5em}
    \dd^n\xii \, p(\xivec)
    \, , \qquad D(\Delta N) \equiv \qty{\xivec: |\Delta N - \Delta N_\xivec | <  \frac{\dd(\Delta N)}{2}} \, ,
\end{equation}
where $\Delta N_\xivec$ is the $\Delta N$ value corresponding to noise given by the vector $\xivec$. The integral is centered around the most probable path, and I call the noise on this path $\mxivec$, a vector with components $\mxii$. However, integrating over the region $D(\Delta N)$ around this path is non-trivial. To obtain a simple analytical estimate, let us assume that the constant-$\Delta N$ hypersurfaces of the $\xii$-space are approximately perpendicular to $\mxii$, so that integrals over the perpendicular directions are trivial. Let us further assume that $\Neff$ is directly proportional to the vector length $|\mxivec|$, that is, $\dd |\mxivec| / \dd (\Delta N) = |\mxivec|/\Delta N$. This gives
\begin{equation} \label{eq:p_approximation}
    p(\Delta N) = \frac{|\mxivec|}{|\Delta N|} \frac{1}{\sqrt{2\pi}} \exp(-\frac{1}{2}|\mxivec|^2) \, ,
\end{equation}
where
\begin{equation} \label{eq:xi_length_on_path}
    |\mxivec|^2 = \int_{N_\ini}^{N_\fin} \frac{(\Neff'-1)^2}{\Pe(\Neff)} \dd N
    = \int_{N_\ini}^{N_\fin - \Delta N} \frac{\Big(\sqrt{1-c\Pe(\Neff)}-1\Big)^2}{\Pe(\Neff)\sqrt{1-c\Pe(\Neff)}} \dd \Neff \, .
\end{equation}
Note that \eqref{eq:xi_length_on_path} only depends on $\Delta N$ and $c$; we don't need to solve \eqref{eq:dneff_path_field_match} explicitly to compute the estimate \eqref{eq:p_approximation}.

For small $\Delta N$, corresponding to a small $c$ and $\Neff \approx N$, \eqref{eq:delta_n_vs_c} expands to give
\begin{equation} \label{eq:delta_n_vs_c_gauss}
    \Delta N \approx \frac{c}{2}\int_{N_\ini}^{N_\fin} \Pe(N)\dd N 
    \, , \qquad
    |\Delta N| \ll 1 \, .
\end{equation}
Expanding \eqref{eq:xi_length_on_path} similarly and plugging in $c$ from \eqref{eq:delta_n_vs_c_gauss} gives
\begin{equation} \label{eq:xi_length_on_path_gauss}
    |\mxivec|^2 \approx \frac{(\Delta N)^2}{
    \int_{N_\ini}^{N_\fin} \Pe(N)\dd N}
    \, , \qquad |\Delta N| \ll 1 \, .
\end{equation}
With this, the approximation \eqref{eq:p_approximation} matches the Gaussian approximation from section~\ref{sec:Gaussian_limit}. The distribution starts to deviate from the Gaussian one when, roughly speaking, $\Delta N$ approaches one.

The integral \eqref{eq:xi_length_on_path} can also be simplified in the limit of large positive $\Delta N$ and thus large $c$, the regime interesting for PBH formation. In this limit, $c\Pe$ in \eqref{eq:dneff_path_field_match} crosses one at some point near its peak. When $\Neff$ approaches this point, its derivative goes to zero and it freezes: the stochastic noise exactly balances out the classical drift. This happens at different $\Neff$ values for different $\Delta N$. If we estimate that $\Neff'$ is one before the transition and jumps sharply to zero, the transition time obeys $N = \Neff = N_c - \Delta N$, and \eqref{eq:xi_length_on_path} yields
\begin{equation} \label{eq:xi_length_in_tail}
    |\mxivec|^2 \approx \int_{N_\fin - \Delta N}^{N_\fin} \frac{\dd N}{\Pe(N_\fin - \Delta N)}
    = \frac{\Delta N}{\Pe(N_\fin - \Delta N)} \, , \qquad \Delta N \gg 1 \, .
\end{equation}
If $\Pe$ is roughly constant at these $N$ (say, near its peak), this produces an exponential tail in \eqref{eq:p_approximation}, with the slope proportional to the inverse of the power spectrum. This is consistent with earlier predictions of exponential tails in stochastic setups \cite{Ezquiaga:2019ftu}.

Such simple estimates are unfortunately not available for large negative $\Delta N$. However, we can deduce the general behavior of $p(\Delta N)$ there compared to the Gaussian estimate from section~\ref{sec:Gaussian_limit}. In that limit, $\Delta N$ got independent contributions of size $\sqrt{\Pe(N)\dd N}\xii$ at every time step. Now, with \eqref{eq:discrete_neff_eom_field_matched}, $\Pe(N)$ is replaced by $\Pe(\Neff)$, where $\Neff > N$ for $\Delta N < 0$. The difference is significant at late times, that is, near $N=N_c$, where $\Pe$ is a decreasing function (see figure~\ref{fig:PRgrid}), and thus $\Pe(\Neff) < \Pe(N)$. To reach the same $\Delta N$, stronger kicks $\xii$ are then needed in the field-matched case to compensate for the smaller $\Pe$, suppressing $p(\Delta N)$ compared to the Gaussian case. The opposite is true for $\Delta N > 0$: stochastic noise pushes the system to stay at larger $\Pe$ values, requiring smaller kicks and enhancing $p(\Delta N)$ compared to the Gaussian estimate. This is consistent with the exponential tails mentioned above, and it also matches the behavior seen in all the numerical examples in \cite{Figueroa:2020jkf, Figueroa:2021zah} and below in section~\ref{sec:numerics}. The behavior is determined by the derivative of $\Pe$ (in practice, $\PR$) at the final coarse-graining scale; if we set $N_c$ to the rising edge of the power spectrum peak in figure~\ref{fig:PRgrid}, the effect would be reversed.

Finally, let me comment on the importance of choosing the coarse-graining parameter $\sigma$. Changing $\sigma$ shifts $\Pphiseff$ and thus $\Pe$ back and forth in $\Neff$, but, assuming the modes are frozen when they reach the coarse-graining scale, does not change its shape. Moreover, to maintain the same final coarse-graining scale, $N_c$ should be changed with $\sigma$ so that the final mode contributing a kick is independent of $\sigma$. Hence, as long as $\epseff_1$ in \eqref{eq:discrete_neff_eom_field_matched} is negligible, the solutions of \eqref{eq:discrete_neff_eom_field_matched} are not sensitive to $\sigma$. This explains the insensitivity of $p(\Delta N)$ to $\sigma$ that was noted in \cite{Figueroa:2021zah}.

These analytical results shed some light on the behavior of $p(\Delta N)$, but to compute it accurately, we must resort to numerics. The results of this section will still be helpful: it turns out that the most efficient way to resolve $p(\Delta N)$ is to compute the volume factor in \eqref{eq:p_to_dn} using the method of importance sampling around the most probable paths $\mxivec$.

\subsection{E-fold matched perturbations}
\label{sec:efold_matching}
For the e-fold matched perturbations, equation~\eqref{eq:discrete_neff_eom} gives
\begin{equation} \label{eq:discrete_neff_eom_efold_matched}
    \dd \Neff = \dd N + \sqrt{\Pe(N,\Neff) \dd N} \, \xii \, , \quad
    \Pe(N,\Neff) \equiv
    \frac{\Pphiseff(N)}{2\Eeff_1(\Neff)} \, ,
    \quad
    \Eeff_1(\Neff) \equiv \frac{\epseff_1(\Neff)}{1-\epseff_1(\Neff)} \, .
\end{equation}
This depends on both $N$ and $\Neff$, complicating the analysis. However, \eqref{eq:discrete_neff_eom_efold_matched} turns out to be the most realistic way to match the perturbations: it correctly reproduces the numerical results of \cite{Figueroa:2020jkf, Figueroa:2021zah} with backreaction between the perturbations and the local background included. To see why, consider the evolution of the perturbations $\dphi_k$ in the post-USR regime where they give their stochastic kicks. As discussed above, the field is there in constant roll with a time-independent $\eps_2$. We have established that the perturbations are frozen, that is, $\R_k = \dphi_k/\sqrt{2\eps_1}$ is a constant, so $\dphi_k \sim \sqrt{\eps_1} \sim a^{\eps_2/2}$. This applies not only on the classical trajectory but also in the presence of stochastic noise: the noise moves $\phi$ back and forth, but it does not change the constant $\eps_2$, and thus it does not change the evolution of $\dphi_k$. In other words, $\Pphi$ takes exactly the same value at $N$ on the classical trajectory and in the full solutions of \eqref{eq:bg_eom}--\eqref{eq:pert_eom} with backreaction included. This was---somewhat accidentally---found out in \cite{Figueroa:2021zah}; we now know how this behavior arises and can exploit it to write down the simplified but identical stochastic process \eqref{eq:discrete_neff_eom_efold_matched}. Note that this logic does not apply outside of constant roll; there one must solve the mode equations \eqref{eq:pert_eom} simultaneously with the background to get accurate results, complicating the computation considerably.

\begin{figure}
    \centering
    \includegraphics{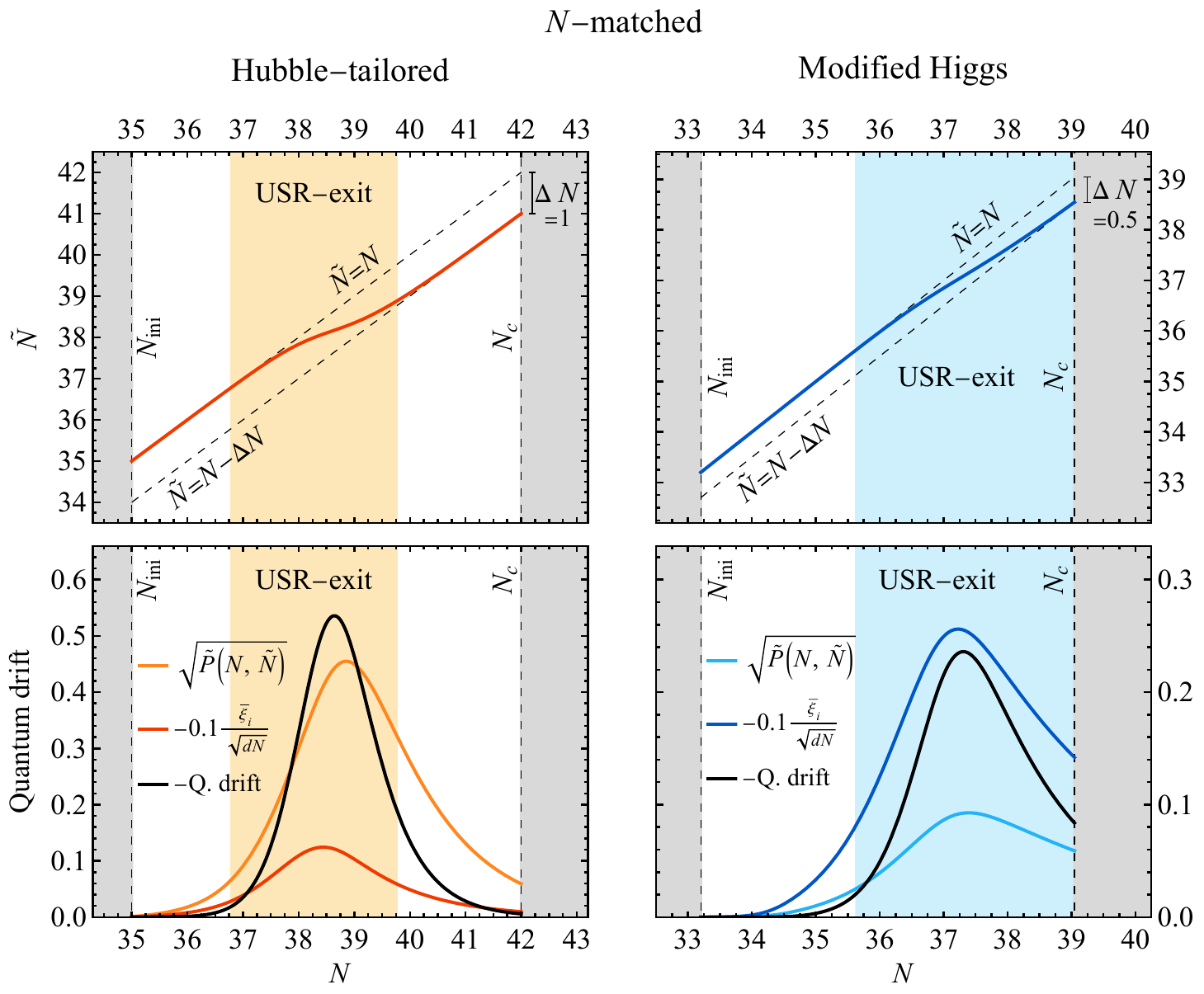}
    \caption{Examples of the most probable paths $\Neff(N)$ in the e-fold matched case, solved from \eqref{eq:neff_path_efold_match}, and the corresponding noises, similarly to figure~\ref{fig:field_matching}.}
    \label{fig:N_matching}
\end{figure}

Equivalently to \eqref{eq:p_xi}--\eqref{eq:neff_path_field_match}, we can derive the most probable paths from the action
\begin{equation} \label{eq:S_xi_efold_match}
    S_\xi =
    -\int_{N_\ini}^{N_\fin} \frac{(\Neff'-1)^2}{2\Pe(N,\Neff)} \dd N \, ,
\end{equation}
giving the equation of motion
\begin{equation} \label{eq:neff_path_efold_match}
    \Neff'' - \frac{\Eeff_1'(\Neff)}{2\Eeff_1(\Neff)}\qty(1 - \Neff'^2) + \frac{\Pphiseff'(N)}{\Pphiseff(N)}\qty(1 - \Neff') = 0 \, .
\end{equation}
This equation has to be solved numerically. A trajectory corresponding to a given $\Delta N$ can be found by a shooting method, varying the initial $\Neff'$. Example solutions are shown in figure~\ref{fig:N_matching}, and they follow the same pattern as in the field-matched case.

The main results from the previous section are still true in the e-fold matched case. The analytical approximation \eqref{eq:p_approximation} can be computed from the solutions of \eqref{eq:neff_path_efold_match} with
\begin{equation} \label{eq:xi_length_on_path_N_matched}
    |\mxivec|^2 =
    \int_{N_\ini}^{N_\fin} \frac{(\Neff'-1)^2}{\Pe(N,\Neff)} \dd N \, .
\end{equation}
Swapping $\Pe(\Neff)$ for $1/\Eeff_1(\Neff)$, the arguments for suppression and enhancement with respect to the Gaussian estimate still apply. Similarly, noting that $\Pphiseff(N)$ is approximately constant during the CR phase, the results are still independent of $\sigma$.

\section{Numerical computations}
\label{sec:numerics}
I solved equations \eqref{eq:discrete_neff_eom_field_matched} and \eqref{eq:discrete_neff_eom_efold_matched} numerically with a {\CC} code a large number of times for both the Hubble-tailored and modified Higgs models and collected statistics on $\Delta N$. The background evolution $\phieff$ and the power spectrum $\Pphiseff$ were computed beforehand to form the functions $\Pe(\Neff)$ and $\Pe(N,\Neff)$ that enter the equations. At each time step, a Gaussian random number $\xii$ was produced using a Mersenne Twister pseudorandom number generator of the {\CC} standard library, and the value of $\Neff$ was updated according to the equation of motion using Euler's method.

The direct solutions of \eqref{eq:discrete_neff_eom_field_matched} and \eqref{eq:discrete_neff_eom_efold_matched} give the probability distribution $p(\Delta N)$ by binning the $\Delta N$ results into bins of width $\dd (\Delta N)$.
If $n_\text{bin}$ is the number of runs in a bin centered around $\Delta N$, then
\begin{equation} \label{eq:p_form_direct_sampling}
    p(\Delta N) = \frac{n_{\text{bin}}}{\dd (\Delta N) n_{\text{tot}}} \, ,
\end{equation}
where $n_\text{tot}$ is the total number of runs. Alternatively, one can employ importance sampling \cite{ImportanceSampling} to resolve $p(\Delta N)$ at a specific $\Delta N$ by introducing a bias to the stochastic noise. The method was first used for stochastic inflation in \cite{Jackson:2022unc}, and it was shown to significantly speed up the computation of the tail of the $\Delta N$ distribution. In \cite{Jackson:2022unc}, a suitable bias was found by trial and error, but we can do better by using the most probable paths from section~\ref{sec:new_formalism}. Let us write the noise as
\begin{equation} \label{eq:xi_decomposition}
    \xii = \mxii + \dxii \, .
\end{equation}
With this change of variables, the integral \eqref{eq:p_to_dn} can be written as
\begin{equation} \label{eq:p_to_dn_biased}
\begin{gathered}
    p(\Delta N)\dd (\Delta N) =
    \hspace{-1em}
    \int\displaylimits_{D(\Delta N)}
    \hspace{-0.5em}
    \dd^n\dxii \, w(\dxivec,\mxivec) p(\dxivec)
    \, ,
    \\
    w(\dxivec,\mxivec) \equiv \exp(-\frac{1}{2}|\mxivec|^2 - \mxivec\cdot\dxivec) \, , \qquad p(\dxivec) \equiv \frac{1}{(2\pi)^{n/2}} \exp(-\frac{1}{2}|\dxivec|^2) \, \, .
\end{gathered}
\end{equation}
In other words, $p(\Delta N)\dd (\Delta N)$ is the expectation value of the function $ w(\dxivec,\mxivec)$ restricted to the bin $D(\Delta N)$ with $\dxii$ as Gaussian random variables with $\expval{\dxii} = 0$, $\expval{\dxii\dxij} = \delta_{ij}$. Drawing random numbers from this distribution and using them to build the full noise \eqref{eq:xi_decomposition}, we can generate multiple runs with \eqref{eq:discrete_neff_eom_field_matched} and \eqref{eq:discrete_neff_eom_efold_matched}, and compute the probability distribution at $\Delta N$ from them as 
\begin{equation} \label{eq:p_form_importance_sampling}
    p(\Delta N) = \frac{\sum_{D(\Delta N)}  w(\dxivec,\mxivec)}{\dd (\Delta N) n_\text{bias}} \, .
\end{equation}
Here $n_\text{bias}$ is the total number of runs generated for the bias $\mxivec$, and the sum is taken over the subset of runs that lie inside the desired bin. Due to the biased sampling, runs in the middle of the bin are sampled most frequently, and \eqref{eq:p_form_direct_sampling} converges fast. By repeating this procedure for multiple bins with different $\Delta N$, corresponding to different biases $\mxivec$ solved as described in section~\ref{sec:new_formalism}, we can resolve the probability distribution very efficiently all the way to its tail.

\begin{figure}
    \centering
    \includegraphics{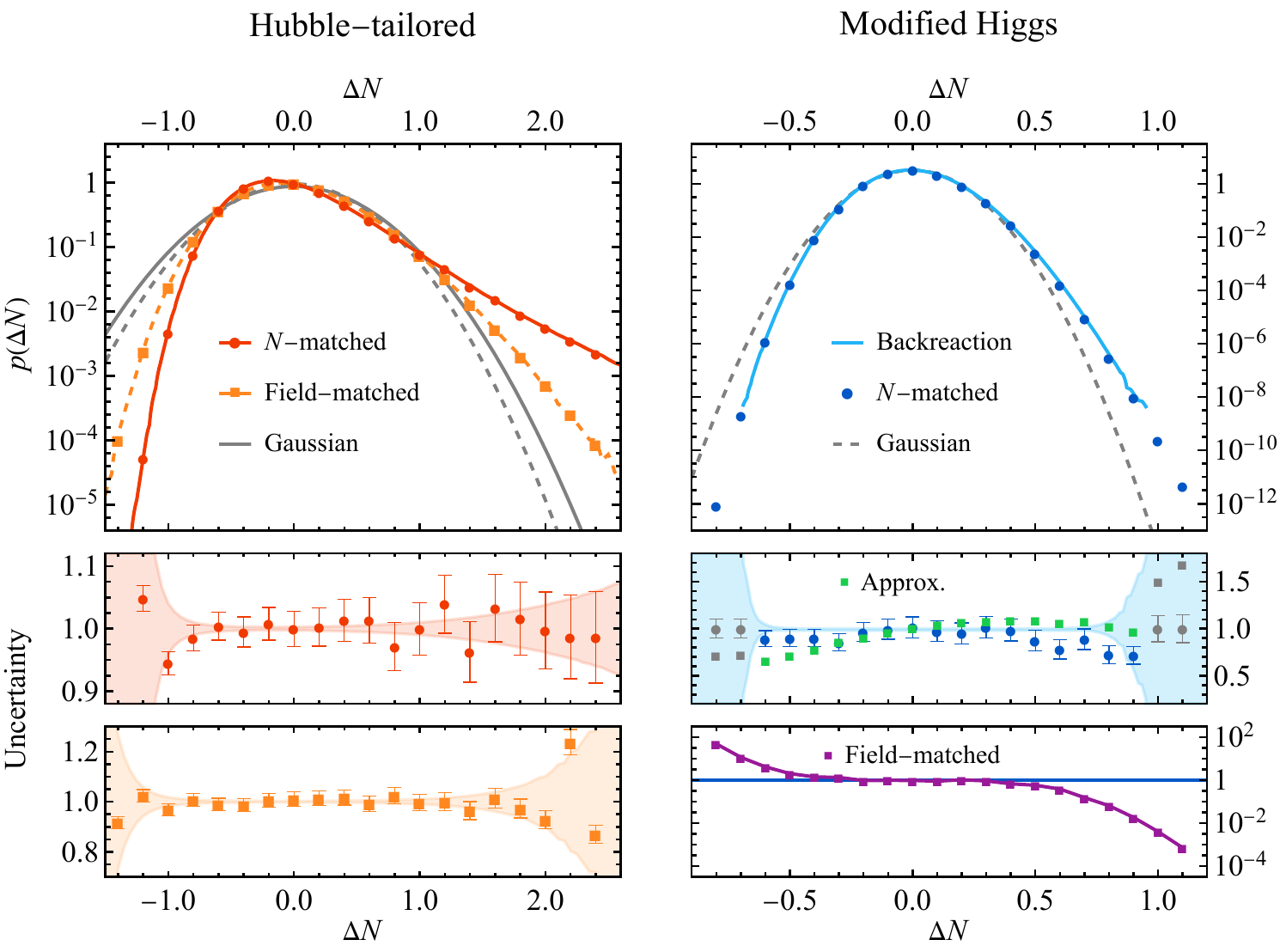}
    \caption{Numerically computed probability distributions.
    \emph{Left:} The Hubble-tailored model. The top panel depicts the $N$-matched direct sampling (red line, interpolation), its Gaussian approximation (gray solid line), and importance sampling (red dots); and the field-matched direct sampling (dashed orange line), its Gaussian approximation (gray dashed line), and importance sampling (orange squares). The lower panels are normalized to the direct sampling value and display the direct sampling $2\sigma$ errors (shaded region) and the importance sampling points with $2\sigma$ error bars, separately for the $N$- and field-matched cases with consistent color coding.
    \emph{Right:} The modified Higgs model. The top panel presents the backreaction computation results from \cite{Figueroa:2021zah} (light blue line), its Gaussian approximation (gray dashed line), and the $N$-matched importance sampling from this paper (blue dots). The mid panel is normalized to the backreaction computation and presents its $2\sigma$ errors (shaded region), the importance sampling points with their $2\sigma$ errors, and the approximation \eqref{eq:p_approximation} (green squares). In regions with no backreaction result (gray marks), normalization follows the importance sampling instead. The bottom panel shows the field-matched, importance-sampled results (purple) relative to the $N$-matched ones.}
    \label{fig:p}
\end{figure}

\paragraph{Hubble-tailored model.} In the Hubble-tailored model, all runs started at $N_\text{ini}=35$ and ended at $N_c =42$. I chose the starting point so that it occurred before the peak in the power spectrum. I placed the end value far in the tail of the power spectrum, see figure~\ref{fig:PRgrid}, to include the stochastic effects over all important scales---increasing $N_c$ did not change the results noticeably.  I divided the interval into $100$ steps of length $0.07$. As convergence tests, lowering the starting time to $N_\text{ini}=33$ and lowering the number of steps to $50$ had no significant effect on the results. Below $50$ steps, the results started to diverge from those obtained here.

Direct sampling included $10^7$ runs, arranged into bins of width $0.05$, with bin mean values running from $-1.5$ to $10.5$. In the field-matched case, the non-empty bins ran from $-1.5$ to $3.5$; in the $N$-matched case, the range was from $-1.3$ to $9.95$. Importance sampling covered $\Delta N$ values from $-1.4$ to $2.4$ at steps of $0.2$, with $10^5$ points generated for each $\Delta N$, and with the same bin width $0.05$ around each value as in direct sampling. 

The top left panel of figure~\ref{fig:p} shows the obtained $\Delta N$ probability distributions around $|\Delta N| \lesssim 2$. This model was tuned to produce strong stochastic effects with highly enhanced tails in $p(\Delta N)$, resolvable up to large $\Delta N$ with a reasonable number of runs even with direct sampling. This is clearly visible in the figure: both the $N$ and field-matched distributions are highly skewed, to the point where the peak of the distribution is shifted from the mean $\Delta N = 0$, and the Gaussian fits (from the mean and variance of the distribution) never approximate $p(\Delta N)$ well. The enhancement of $p(\Delta N)$ for $\Delta N > 0$ and the suppression of $p(\Delta N)$ for $\Delta N < 0$, explained in section~\ref{sec:new_formalism}, are clearly visible. Despite the skewness, the mean run still matches the classical one with no noise, within numerical accuracy. The skewness is stronger for the $N$-matched distribution, and the difference is significant for large $\Delta N$.

As the lower left panels of figure~\ref{fig:p} show, the importance-sampled results match the directly sampled distributions well. Appendix~\ref{sec:errors} explains how the uncertainties were computed. If needed, the uncertainty can be decreased by increasing the number of generated runs. The analytical approximation \eqref{eq:p_approximation}, not plotted, differs from the numerical result by a factor of $0.44$ to $1.8$ in the field-matched case and $0.14$ to $3.7$ in the $N$-matched case, with the factor decreasing with an increasing $\Delta N$. The errors are relatively large; presumably, the strong stochastic kicks make the volume factor in \eqref{eq:p_to_dn} important and not well captured by the approximation.

\begin{figure}
    \centering
    \includegraphics{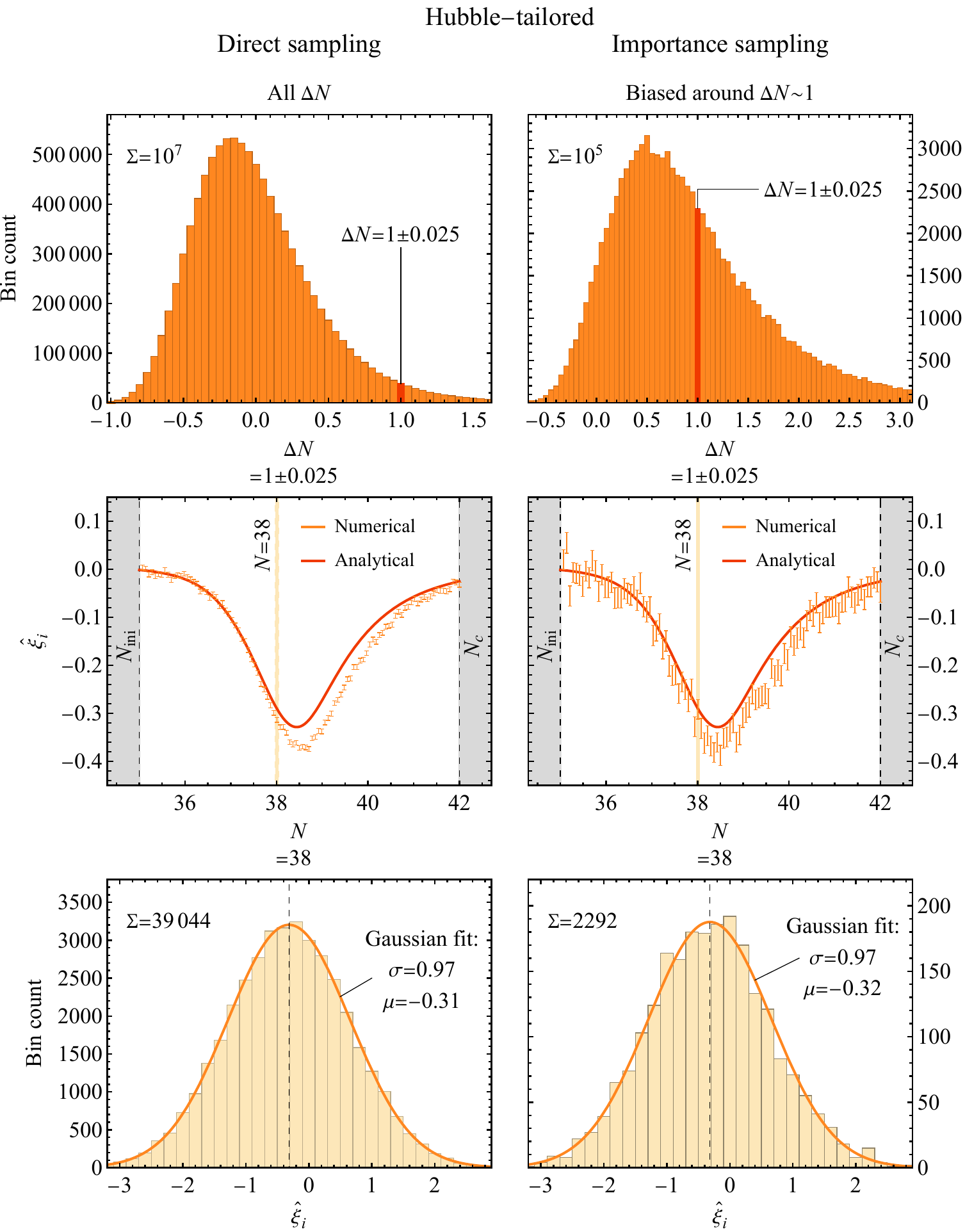}
    \caption{Comparison between the direct and importance samplings in the Hubble-tailored model. The top panels present the full directly sampled results and the importance-sampled results generated for a particular bias. The middle panels show the noise of the most probable path (`Analytical') and the realized noise and its $1\sigma$ errors (`Numerical') step by step in the highlighted bin. The bottom panels display the spread of the noise at a particular time step in the highlighted bin. Both spreads are consistent with a Gaussian distribution. `$\Sigma$' gives the total number count in a histogram.}
    \label{fig:toy_sampling}
\end{figure}

Figure~\ref{fig:toy_sampling} provides a deep dive into the statistics of the stochastic kicks in the $N$-matched case, comparing the direct and importance sampling methods. In direct sampling, $0.39\%$ of the $10^7$ total runs hit the example bin around $\Delta N = 1$. In importance sampling, $2.3\%$ of the $10^5$ runs generated with the $\Delta N = 1$ bias hit the bin. The mean values of the noises $\xii$ in this bin follow the most probable path from section~\ref{sec:new_formalism} adequately, but deviate for large $N$---again, I assume the volume factor plays an important role and introduces corrections to the action \eqref{eq:S_xi} that are hard to capture analytically. The used bias $\mxivec$ is thus not ideal, but $\xii$ do cluster around the same path in both the direct and importance-sampled cases, so this seems to introduce no systematic error in the sampling. Appendix~\ref{sec:errors} discusses the computation of the noise error bars. Zooming in to a specific time step at $N=38$, we see that the $\xii$ values there follow a Gaussian distribution with unit norm, justifying the choice of bias \eqref{eq:xi_decomposition} with $\expval{\dxii^2} = 1$.

All in all, the Hubble-tailored model demonstrates the usefulness of the constrained formalism of section~\ref{sec:new_formalism} and the utility of importance sampling. I then put these techniques to a real test in the modified Higgs case, where stochastic effects are weaker and collecting statistics for large $\Delta N$ is more challenging.

\paragraph{Modified Higgs.} For the modified Higgs case, the runs started at $N_\text{ini}=33.2$ and ended at $N_c=39.047$. The endpoint matches that used in \cite{Figueroa:2020jkf, Figueroa:2021zah}: the last mode to give a kick is the one that exits the Hubble radius at the end of USR, see figure~\ref{fig:PRgrid}. I divided the interval into $100$ steps, yielding a step length of $\dd N =0.05847$. Lowering the starting time did not significantly change the results, nor did increasing the number of time steps to $1000$.

\begin{figure}
    \centering
    \includegraphics{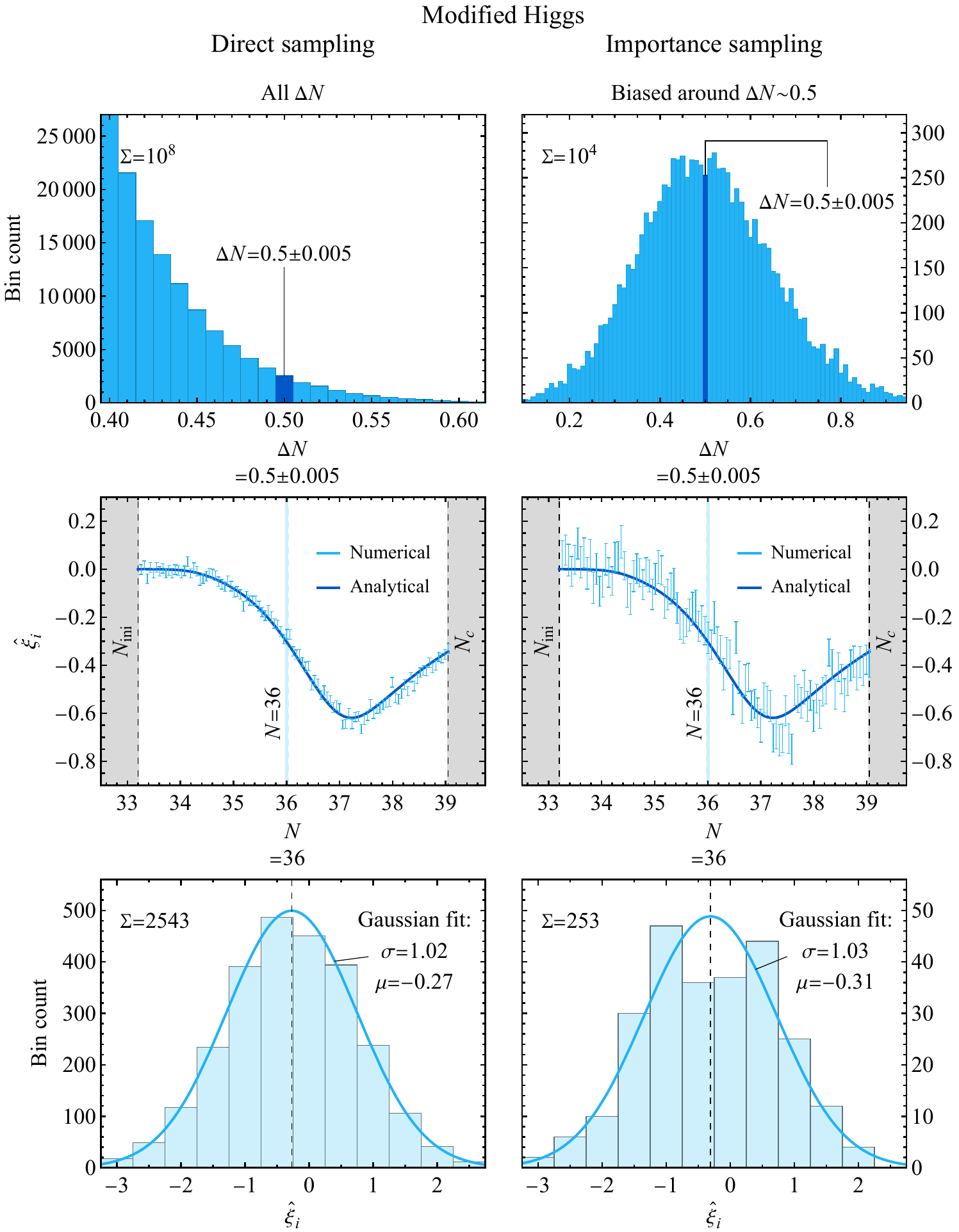}
    \caption{Comparison between the direct and importance samplings in the modified Higgs model, similarly to figure~\ref{fig:toy_sampling}. The $\xii$ spreads in the bottom panels are consistent with Gaussian distributions.}
    \label{fig:higgs_sampling}
\end{figure}

In the modified Higgs case, I performed importance sampling around $\Delta N$ values running from $-1$ to $1.5$ in steps of $0.1$. For each $\Delta N$, I generated $10^4$ runs to compute $p(\Delta N)$ in a bin of width $0.01$. I mainly compared the importance-sampled results to earlier numerical results with backreaction from \cite{Figueroa:2021zah}, with $1024\times 10^8$ runs in bins of width $1/64 \approx 0.016$ running from $-0.69$ to $0.95$. For the detailed comparison of figure~\ref{fig:higgs_sampling}, I also performed $10^8$ $N$-matched runs with direct sampling, producing again data in bins of width $0.01$, running from $-0.61$ to $0.79$.

The right panels of figure~\ref{fig:p} show the results for $p(\Delta N)$. The stochastic kicks are milder than in the Hubble-tailored model: the $\Delta N$ distribution matches the Gaussian approximation near its peak, and the non-Gaussian tails are less pronounced. As anticipated, the $N$-matched importance-sampled results line up with the backreaction computation of \cite{Figueroa:2021zah}. The difference between these two is of order $10\%$, mostly within the statistical uncertainty, although there appears to be a small systematic bias suppressing the importance-sampled results for large $\Delta N$. Presumably, this difference originates from differences in binning and numerical techniques and small violations of the assumptions of perfect freezing and squeezing made in section~\ref{sec:new_formalism}. The difference is not significant for estimating PBH abundances. The field-matched case differs significantly from the $N$-matched one, again displaying weaker skewness. On the other hand, the analytical approximation \eqref{eq:p_approximation} of the $N$-matched case yields good results with the correct order of magnitude, though the error is growing towards large $\Delta N$.

Figure~\ref{fig:higgs_sampling} compares the importance-sampled $N$-matched results to directly sampled ones, similarly to figure~\ref{fig:toy_sampling}. Of the directly sampled runs, $0.0025\%$ lie in the example bin at $\Delta N = 0.5$. In the importance-sampled case, $2.5\%$ of the runs generated for the bias hit the bin. This is of the same order as in the Hubble-tailored model; the distribution here is narrower (due to lower $\PR$), which boosts the ratio, but the chosen bin width is narrower too. Increasing the bin width would lead to a higher hit rate and a faster convergence of $p(\Delta N)$, with some loss of resolution in the $\Delta N$ direction. Again, the $\xii$ distribution at a fixed time step is Gaussian with $\expval{\dxii^2} = 1$. However, now $\xii$ follows the most probable path $\mxii$ well: for such a `realistic' example with lower stochastic noise, the constrained formalism of section~\ref{sec:new_formalism} is very accurate.

\begin{table}
\begin{center}
\begin{tabular}{l C{2cm} C{2cm} C{2cm} C{2cm}}
\toprule
& \multicolumn{2}{c}{\textbf{Hubble-tailored}} & \multicolumn{2}{c}{\textbf{Modified Higgs}} \\
& Runs & CPU time & Runs & CPU time \\
\midrule
\textbf{Field-matched} &&&& \\
Direct & $10^7$ & $72\,\text{s}$ & \multicolumn{2}{c}{\rule[0.8ex]{2em}{0.2pt}} \\
Importance & $20\times 10^5$ & $15\,\text{s}$ & $26 \times 10^4$ & $3\,\text{s}$ \\
\midrule
\textbf{\boldmath $N$-matched} &&&& \\
Direct & $10^7$ & $74\,\text{s}$ & $10^8$ & $688\,\text{s}$ \\
Importance & $20\times 10^5$ & $16\,\text{s}$ & $26 \times 10^4$ & $2\,\text{s}$ \\
\midrule
\textbf{Backreaction} \cite{Figueroa:2021zah} & \multicolumn{2}{c}{\rule[0.7ex]{2em}{0.2pt}}  & $1.6 \times 10^{9}$ & $\sim10^6\,\text{h}$ \\
\bottomrule
\end{tabular}
\end{center}
\caption{A summary of the number of stochastic runs and the corresponding time of computation for each process. With the method of this paper, one run took approximately $7\times 10^{-6}$ seconds, with some overhead from setting up the most probable paths for importance sampling and saving the results into files.}
\label{tab:run_stats}
\end{table}

Table \ref{tab:run_stats} shows the number of runs and the CPU time used for the different data sets on a $2.3 \, \text{GHz}$, $6$ core laptop. We see that the time saved by importance sampling is significant. Moreover, with importance sampling, the time cost to compute a point in the tail of $p(\Delta N)$ is almost independent of $\Delta N$, while for direct sampling, it increases exponentially in $\Delta N$. Particularly impressive is the time saved between the directly sampled modified Higgs case with backreaction from \cite{Figueroa:2020jkf, Figueroa:2021zah}, which took of order one million CPU hours, and the importance-sampled computation here, completed in two seconds---an improvement of factor $10^9$, with the importance-sampled distribution extending significantly farther into the tail. The bottleneck in all computations was random number generation for the noise, which took approximately $70\%$ of the running time. The CPU time can be cut down even more, to a negligible amount, by using the analytical approximation \eqref{eq:p_approximation}, although the quality of the approximation varies, as explained above.

\clearpage

\clearpage

\section{Discussion}
\label{sec:discussion}

The numerical examples show that the constrained stochastic inflation formalism of section~\ref{sec:new_formalism} is useful for quick but accurate computation of the $p(\Delta N)$ distribution for large $\Delta N$, especially in its importance sampling form. Importance sampling was used earlier for stochastic inflation in the slow-roll limit in \cite{Jackson:2022unc}. There, the authors chose the optimal bias by trial and error, while the constrained formalism lets us compute the most probable path semi-analytically and use this as an optimized bias. The authors of \cite{Jackson:2022unc} used a handful of different biases to estimate $p(\Delta N)$ over a wide range of $\Delta N$---since the optimal bias for a particular $\Delta N$ is easy to compute in the method of this paper, I instead advocate doing this separately for each desired $\Delta N$ point, producing one bin per bias, as demonstrated in section~\ref{sec:numerics}.

In \cite{Jackson:2022unc}, one bin of importance-sampled data contained weights $w$ of vastly different magnitudes, and the largest ones dominated in their version of the sum \eqref{eq:p_form_importance_sampling}. To fix this, they fitted a lognormal estimator to the weight distribution to approximate the sum. This problem did not appear in the numerics of the current paper, done with the bias \eqref{eq:xi_decomposition}: all runs near the biased $\Delta N$ had weights of the same order. I also ran tests where the variance of $\dxii$ in \eqref{eq:xi_decomposition} was not equal to one, and these produced the weight problem of \cite{Jackson:2022unc}, together with slower convergence and a biased $p(\Delta N)$ distribution. This suggests that the bias \eqref{eq:xi_decomposition} with $\expval{\dxivec^2}=1$ is indeed an optimal one, or very close to it. 

The main usage for $p(\Delta N)$ for large $\Delta N$ is the computation of PBH abundances. To this day, a Gaussian approximation computed from $\PR$ is often employed in the literature due to its simplicity, even though it fails for the relevant perturbations of strength $\Delta N \sim 1$ in typical models, as we saw above.  The results of this paper offer an alternative, easy-to-use but more accurate method: the analytical approximation \eqref{eq:p_approximation}. For this, one only needs to solve the $N$-matched differential equation \eqref{eq:neff_path_efold_match} with initial conditions that produce the right $\Delta N$ and plug the solution into \eqref{eq:xi_length_on_path_N_matched}. Only a few $p(\Delta N)$ points are needed to resolve the tail around the wanted perturbation strength.

The approximation also transitions smoothly into the standard Gaussian one in the small $\Delta N$ limit. In general, the constrained formalism of section \ref{sec:new_formalism} makes the role of $\PR$ and the Gaussian approximation very transparent in the computation of $p(\Delta N)$, and can also shed light on when the Gaussian approximation applies and when it breaks.

The choice of final boundary conditions in this paper and in \cite{Figueroa:2020jkf, Figueroa:2021zah} differs from that of many other studies, which employ the first passage time formalism \cite{Vennin:2015hra, Pattison:2017mbe, Pinol:2018euk, Ezquiaga:2019ftu, Prokopec:2019srf, Ando:2020fjm, Rigopoulos:2021nhv, Pattison:2021oen, Tada:2021zzj, Ahmadi:2022lsm, Mahbub:2022osb, Animali:2022otk, Jackson:2022unc}. There the stochastic kicks continue all the way to the end of the computation, which terminates when the field first crosses the final hypersurface at $\phi=\phi_\text{final}$. Changing $\phi_\text{final}$ can then probe different perturbation scales, similarly to different choices of the final kick time $N_c$ in this paper's formalism. The advantage of the first passage time formalism is that highly developed tools exist to solve the stochastic system semi-analytically. On the other hand, the formalism of this paper makes more direct contact with the coarse-graining scale and the physical interpretation of the perturbations.

However, if the coarse-graining scale of interest is shifted from the peak in $\PR$, then the two formalisms should produce essentially the same results. This is true to an extent in our modified Higgs model and more clearly in the Hubble-tailored model (see figure~\ref{fig:PRgrid}). In this case, any kicks after $N_c$ would be weak and subdominant. It then makes no difference to evolve from $N_c$ to the final $\phi=\phi_\text{final}$ hypersurface with or without the kicks, matching the first passage time formalism and our constrained formalism, respectively.

Finally, let us recap the assumptions used to arrive at the constrained stochastic method. For the stochastic kicks to align with the classical trajectory, we need the perturbations to be frozen (and thus also squeezed) when they arrive at the coarse-graining scale---this requires a small enough coarse-graining parameter $\sigma$ so that the enhanced USR perturbations only give their kicks after the end of USR. As we have seen, this is not a problem for coarse-graining scales near the peak of the power spectrum $\PR$, but one needs to be careful when probing longer scales that exit the Hubble radius at the beginning of USR or slightly earlier. On the other hand, we also saw that as long as $\sigma$ is small enough, its exact value is not very relevant for the results.

When employing the most accurate $N$-matched variation of the constrained method, I also assumed that the system is in CR with a constant $\eps_2$ when the most important scales deliver their stochastic kicks. This guarantees that the pre-computed perturbations are still valid in the stochastic background. Again, a small enough $\sigma$ guarantees this, if the USR is indeed followed by a long enough CR phase. Note, though, that high enough $\Delta N$ will always push the system out of CR and back into the USR phase; for such extremely high $\Delta N$, the method presented here cannot be trusted.

\section{Conclusions}
\label{sec:conclusions}
In this paper, I sought to gain an analytical understanding of the results of \cite{Figueroa:2020jkf, Figueroa:2021zah}, where the equations of stochastic inflation were solved numerically in PBH-producing single-field models, including backreaction between the coarse-grained variables and the short-wavelength perturbations. I showed that since the perturbations are frozen and squeezed when they give stochastic kicks, they keep the system on the original classical track in phase space, only moving it back and forth along this track. This was postulated earlier in \cite{Prokopec:2019srf, Cruces:2021iwq, Rigopoulos:2021nhv, Cruces:2022imf} based on the momentum constraint of Einstein equations; I showed it follows from the perturbation dynamics alone, as long as the coarse-graining scale is far enough removed from the Hubble scale and the perturbations are solved accurately, beyond the de Sitter approximation.

With this insight, I reformulated the problem into \emph{constrained stochastic inflation}, where the stochastic degree of freedom is the number of e-folds along the classical trajectory. With this, it is easy to compute curvature perturbations through the $\Delta N$ formalism. Pre-computed short-wavelength perturbations determine the strength of the stochastic noise. I studied two ways to match the short-wavelength perturbations to the stochastic evolution, by field value and by e-folds, and found the second option to mimic the approach of \cite{Figueroa:2020jkf, Figueroa:2021zah}. This works because the field is in constant-roll inflation when it experiences the strongest stochastic kicks, and constant-roll dynamics erase all backreaction between the short and long-wavelength perturbations. In absence of backeaction, all non-Gaussianity originates from the non-linear background dynamics and the $\Delta N$ formalism. I presented a way to compute the most probable noise configuration for a given $\Delta N$ and derived an analytical approximation for the probability distribution $p(\Delta N)$.

I solved the stochastic equations numerically in two example models, one built by hand to produce large perturbations and the other picked from \cite{Figueroa:2020jkf, Figueroa:2021zah} to allow a comparison of the results. The fastest way to compute the probability distribution up to its tail turned out to be importance sampling around the most probable paths. This produced results compatible with \cite{Figueroa:2020jkf, Figueroa:2021zah} at the $10\%$ level, very accurate considering the exponential sensitivity to $\Delta N$, but with a considerable saving in computational cost---one million CPU hours in \cite{Figueroa:2020jkf, Figueroa:2021zah} versus a few seconds with the new method. The analytical approximation was almost as accurate with next to no computational cost.

When the study of PBHs from inflation evolved, it was realized that the slow-roll approximation breaks down in PBH-producing models with a feature in the potential. The perturbation power spectrum then has to be solved numerically from the Sasaki--Mukhanov equation. In a similar fashion, the improved stochastic computations of this process should move beyond the de Sitter approximation and use short-wavelength perturbations that are solved numerically.

This paper presents one way to perform such computations in a well-motivated and computationally feasible way. It is suitable for tuning parameters accurately to produce a desired PBH abundance from the tail, a feat that has thus far been practically out of reach for computations with this level of rigor. Using the analytical approximation, in particular, is no more expensive than numerically computing the power spectrum $\PR$ over a number of modes, already standard practice in Gaussian PBH studies that go beyond the SR approximation.

\acknowledgments
I thank Daniel Figueroa, Sami Raatikainen, and Syksy Räsänen for collaboration on past stochastic inflation projects, which this paper builds on. I also thank Archie Cable, Joe Jackson, Vincent Vennin, and Ashley Wilkins for discussions. This work was supported by the Estonian Research Council grant PRG1055 and by the EU through the European Regional Development Fund CoE program TK133 ``The
Dark Side of the Universe.''

\appendix

\section{Hubble-tailored model}
\label{sec:eff_model}

To build the Hubble-tailored model, I postulate the form of the first slow-roll parameter:
\begin{equation} \label{eq:eff_eps1_construction}
\begin{aligned}
    \eps_1 &= \eps_{1,\text{top}} \times g^2_{\text{USR-CR}} \times g^2_{\text{cut}} \times g^2_{\text{SR}} \, ,
    \\
    g_{\text{USR-CR}} &\equiv e^{-\frac{3}{2}(N-N_1)}\frac{\cosh[\lambda(N_2-N)]}{\cosh[\lambda(N_2-N_1)]} \, ,
    \\
    g_\text{cut} &\equiv \qty[\frac{2}{1+e^{-\theta_\text{cut}(\lambda + 3/2)(N-N_1)}}]^{1/\theta_\text{cut}} \, ,
    \\
    g_{\text{SR}} &\equiv \frac{1+\sqrt{\alpha}\qty(\theta_\text{SR}^{-1}\ln 2)^{\beta/2}}{1+\sqrt{\alpha}\qty(\theta_\text{SR}^{-1}\ln\hspace{-0.1cm}\qty[1+e^{-\theta_\text{SR}(N-N_1)}])^{\beta/2}} \, .
\end{aligned}
\end{equation}
Here, the factor $g_{\text{USR-CR}}$ determines the behavior of $\eps_1$ in the USR and the following CR phase in a way compatible with the Wands duality, as discussed recently in \cite{Karam:2022nym}. The parameter $\lambda$ sets the duration of this phase (ending with $\eps_1=1$) and the second slow-roll parameter there, and the times $N_1$ and $N_2$ roughly determine the beginning and end of USR. The length of USR, $N_2-N_1$, controls the height of the ensuing power spectrum peak. The factor $g^2_{\text{cut}}$ tames the USR-CR behavior at early times, and the factor $g^2_{\text{SR}}$ introduces a gentler, plateau-like SR behavior there, modifiable through $\alpha$ and $\beta$. The constants $\theta_\text{cut}$ and $\theta_\text{SR}$ determine the sharpness of the transition from SR to USR. There is a local maximum in $\eps_1$ around this transition; $\eps_{1,\text{top}}$ is the approximate value of $\eps_1$ there.

The behavior of \eqref{eq:eff_eps1_construction} in the different phases can be summarized as
\begin{equation} \label{eq:eff_eps_behavior}
\begin{aligned}
    &N < N_1: \quad && g^2_{\text{USR-CR}} \times g^2_{\text{cut}} \sim \text{const.} \\
    &&& \eps_1 \sim g^2_{\text{SR}} \sim \frac{1}{\alpha(N_1-N)^\beta} \\
    &N = N_1: && \eps_1 = \eps_{1,\text{top}} \\
    & N > N1: && g^2_{\text{cut}} \, , \, g^2_{\text{SR}} \sim \text{const.} \\
    &&& N < N_2: \, \eps_1 \sim e^{-(3+2\lambda)N} \, , \, \eps_2 \approx -3-2\lambda \\
    &&& N > N_2: \, \eps_1 \sim e^{-(3-2\lambda)N} \, , \, \eps_2 \approx -3+2\lambda \, .
\end{aligned}
\end{equation}
The parameter values used in this paper are
\begin{equation} \label{eq:eff_model_parameter_values}
\begin{gathered}
    \eps_{1,\text{top}} = 0.01 \, , \quad
    N_1 = 32 \, , \quad
    N_2 = 35.04 \, , \quad
    \lambda = 2.308 \, ,
    \\
    \alpha = 50 \, , \quad
    \beta = 1.28 \, , \quad
    \theta_\text{cut} = 1 \, , \quad
    \theta_\text{SR} = 5 \, .
\end{gathered}
\end{equation}
The corresponding $\eps_1$ and $\eps_2$ are plotted in figure~\ref{fig:Vepsgrid}. The number of e-folds $N$ is computed from the CMB pivot scale, placed so that inflation ends 50 e-folds after CMB. The CMB observables \eqref{eq:CMB} are
\begin{equation} \label{eq:eff_model_CMB}
    n_s \approx 0.960 \, , \qquad r \approx 0.0013 \, ,
\end{equation}
compatible with the observations. The model produces a peak in the curvature power spectrum 16 e-folds before the end of inflation, mimicking the behavior of the modified Higgs model but with stronger perturbations and, thus, stronger stochastic effects.
    
The form of $\eps_1(N)$ fixes the Hubble parameter up to its normalization through \eqref{eq:SR_parameters}, $\eps_1 = -\partial_N \ln H$---hence the moniker `Hubble-tailored model.' The normalization also normalizes the perturbations; I fix it to produce the correct CMB power spectrum, $A_s = H^2/(8\pi^2\eps_1) \approx 2.1\times10^{-9}$. The power spectrum peak then reaches $\PR(k_\text{peak}) \approx 0.10$.

We can further solve the classical $\phi(N)$ from \eqref{eq:SR_parameters}, $\eps_1 = (\partial_N \phi)^2/2$. The potential is given by $V=(3-\eps_1)H^2$. With $\phi(N)$ and $V(\phi(N))$ known, we can numerically solve $V(\phi)$, depicted in figure~\ref{fig:Vepsgrid}.

\section{Estimating statistical errors}
\label{sec:errors}
In \eqref{eq:p_form_direct_sampling}, $p(\Delta N)$ is computed from the number of observations in a bin and can be written as the expectation value of a function $f_\text{bin}$ as
\begin{equation} \label{eq:p_form_direct_sampling_app}
    p(\Delta N) = \expval{f_\text{bin}}  \, , \qquad
    f_\text{bin} =
    \begin{cases}
    \frac{1}{\dd(\Delta N)} & \text{if run in bin,}\\
    0 & \text{otherwise.}
    \end{cases}
\end{equation}
The expectation value is taken over all the runs, and the value of $f_\text{bin}$ for a run depends on whether the run hits or misses the bin. The one sigma uncertainty of $p(\Delta N)$ can now be computed as the standard error of the mean,
\begin{equation} \label{eq:p_sigma_direct}
    \sigma_p = \sqrt{\frac{\expval{f_\text{bin}^2} - \expval{f_\text{bin}}^2}{n_\text{tot}}} \, ,
    \qquad
    \expval{f_\text{bin}} = \dd(\Delta N)\expval{f_\text{bin}^2} = \frac{n_\text{bin}}{\dd(\Delta N)n_\text{tot}} \, .
\end{equation}
Similarly, \eqref{eq:p_form_importance_sampling} can be written as
\begin{equation} \label{eq:p_form_importance_sampling_app}
    p(\Delta N) = \expval{f_{w,\text{bin}}}  \, , \qquad
    f_{w,\text{bin}} =
    \begin{cases}
    \frac{w}{\dd(\Delta N)} & \text{if run in bin,}\\
    0 & \text{otherwise.}
    \end{cases}
\end{equation}
Now, $f_{w,\text{bin}}$ varies inside the bin; the weight $w=w(\dxivec,\mxivec)$ for a run is defined in \eqref{eq:p_to_dn_biased}. The one sigma uncertainty becomes
\begin{equation} \label{eq:p_sigma_importance}
    \sigma_p = \sqrt{\frac{\langle f_{w,\text{bin}}^2 \rangle - \expval{f_{w,\text{bin}}}^2}{n_\text{bias}}} \, ,
    \quad
    \expval{f_{w,\text{bin}}} = \frac{\sum_{D(\Delta N)} w}{\dd(\Delta N) n_\text{bias}} \, ,
    \quad
    \expval{f_{w,\text{bin}}^2} = \frac{\sum_{D(\Delta N)} w^2}{[\dd(\Delta N)]^2 n_\text{bias}} \, .
\end{equation}
For the step-wise noise averages in the middle rows of figures \ref{fig:toy_sampling} and \ref{fig:higgs_sampling}, the average is computed as the weighted mean over all the runs in the bin, with weights equal to the run weight $w$. The average and its error follow \cite{data-analysis-toolkits-2020}
\begin{equation} \label{eq:noise_average}
    \expval{\xii} = \frac{\sum_j w_j \xiij}{\sum_j w_j} \, ,
    \quad
    \expval{\xii^2} = \frac{\sum_j w_j \xiij^2}{\sum_j w_j} \, ,
    \quad
    \sigma_{\xii} = \sqrt{\frac{\expval{\xii^2} - \expval{\xii}^2}{n_\text{eff}-1}} \, ,
    \quad
    n_\text{eff} = \frac{\qty(\sum_j w_j)^2}{\sum_j w_j^2} \, ,
\end{equation}
where $j$ runs over all the runs in the bin, and $\xiij$ is the noise of the $i$th time step in the $j$th run.

\bibliographystyle{JHEP}
\bibliography{stoc}

\end{document}